\documentclass[reprint,superscriptaddress,amsmath,amssymb,aps,prl]{revtex4-1}
\pdfoutput=1

\usepackage{graphicx}
\usepackage{dcolumn}
\usepackage{bm}
\usepackage{tabularx}
\usepackage{array}
\usepackage{amsmath}
\usepackage{fancyhdr}
\usepackage{amsfonts}
\usepackage{hyperref}
\usepackage[mathlines]{lineno}

\usepackage{soul}

\begin{document}

\newcolumntype{s}{>{\hsize=.2\hsize}X} \newcolumntype{b}{>{\hsize=.6\hsize}X}

\mathchardef\mslash="202F

\makeatletter
\newcommand*\@dblLabelI {}
\newcommand*\@dblLabelII {}
\newcommand*\@dblequationAux {}

\def\@dblequationAux #1,#2,%
    {\def\@dblLabelI{\label{#1}}\def\@dblLabelII{\label{#2}}}

\newcommand*{\doubleequation}[3][]{%
    \par\vskip\abovedisplayskip\noindent
    \if\relax\detokenize{#1}\relax
       \let\@dblLabelI\@empty
       \let\@dblLabelII\@empty
    \else 
       \@dblequationAux #1,%
    \fi
    \makebox[0.5\linewidth-1.5em]{%
     \hspace{\stretch2}%
     \makebox[0pt]{$\displaystyle #2$}%
     \hspace{\stretch1}%
    }%
    \makebox[0.5\linewidth-1.5em]{%
     \hspace{\stretch1}%
     \makebox[0pt]{$\displaystyle #3$}%
     \hspace{\stretch2}%
    }%
    \makebox[3em][r]{(%
  \refstepcounter{equation}\theequation\@dblLabelI, 
  \refstepcounter{equation}\theequation\@dblLabelII)}%
  \par\vskip\belowdisplayskip
}
\makeatother

\preprint{APS/123-QED}

\title{Dependence of phase behavior and surface tension on particle stiffness for active Brownian particles$^\dag$}

\author{Nicholas Lauersdorf}
\affiliation{%
 Department of Applied Physical Sciences, The University of North Carolina at Chapel Hill, Chapel Hill, NC, USA
}%

\author{Thomas Kolb}%
\affiliation{%
 Department of Chemistry, The University of North Carolina at Chapel Hill, Chapel Hill, NC, USA
}%

\author{Moslem Moradi}
\affiliation{%
 Department of Applied Physical Sciences, The University of North Carolina at Chapel Hill, Chapel Hill, NC, USA
}%

\author{Ehssan Nazockdast}
\affiliation{%
 Department of Applied Physical Sciences, The University of North Carolina at Chapel Hill, Chapel Hill, NC, USA
}%

\author{Daphne Klotsa} 
\email{dklotsa@email.unc.edu}
\affiliation{%
Department of Applied Physical Sciences, The University of North Carolina at Chapel Hill, Chapel Hill, NC, USA
}%

\begin{abstract}
We study quasi two-dimensional, monodisperse systems of active Brownian particles (ABPs) for a range of activities, stiffnesses, and densities. We develop a microscopic, analytical method for predicting the dense phase structure formed after motility-induced phase separation (MIPS) has occurred, including the dense cluster's area fraction, interparticle pressure, and radius. Our predictions are in good agreement with our Brownian dynamics simulations. We, then, derive a continuum model to investigate the relationship between the predicted interparticle pressure, the swim pressure, and the macroscopic pressure in the momentum equation. We find that formulating the point-wise macroscopic pressure as the interparticle pressure and modeling the particle activity through a spatially variant body force --as opposed to a volume-averaged swim pressure-- results in consistent predictions of pressure in both the continuum model and the microscopic theory. This formulation of pressure also results in nearly zero surface tension for the phase separated domains, irrespective of activity, stiffness, and area fraction. Furthermore, using Brownian dynamics simulations and our continuum model, we showed that both the interface width and surface tension, are intrinsic characteristics of the system. On the other hand, if we were to exclude the body force induced by activity, we find that the resulting surface tension values are linearly dependent on the size of the simulation, in contrast to the statistical mechanical definition of surface tension. 
\end{abstract}

\maketitle


\section{\S1: Introduction}

Active-matter systems consist of ``active'' components (e.g. self-propelled nanorods, molecular motors) that locally consume energy to move, exert forces or perform chemical reactions, thus being inherently out of equilibrium. Properties of active matter such as adaptation, responsiveness, and self-healing may enable the development of novel materials and technologies \cite{Ghosh2013, Palacci2010, VanDerMeer2016, Ghosh2020, Patterson2010, Brambilla2013, Gao2013, Orozco2014, Li2017, Li2014, Perez2014, Chen2015}. However, to develop these next-generation technologies, a deeper theoretical understanding and description of active systems is needed. In the past two decades both mechanical and thermodynamic approaches, predicated on our understanding of equilibrium matter, have provided great insight towards an understanding of active systems, which are out of equilibrium and violate detailed balance~\cite{Cates2012DiffusivePhysics,Marchetti2016MinimalMatter,Battle2016BrokenSystems,Fodor2016HowMatter,Stenhammar2014, Kolb2020ActiveSpheres}. However, in its current state, active matter has no complete theory, no ``real gas model'' which can predict emergent behavior based on system parameters.

Simulations are a useful platform for testing active-matter theories by allowing the calculation of properties inaccessible or difficult to obtain experimentally, as well as the investigation of a broad parameter space. Here, we focus on the active Brownian particle (ABP), a model subject to the overdamped Langevin equations of motion in which a particle propels itself at an intrinsic speed while rotating randomly in time \cite{Schimansky-Geier1995, Redner2013, Cates2015a}. One of the most surprising and interesting behaviors observed with the ABP model is motility-induced phase separation (MIPS), where the system undergoes a first-order phase transition into dense and dilute (gas-like) phases induced by the activity of the particles in absence of an attractive potential~\cite{Redner2013, Cates2015a}. 

Though it is mainly activity and density that have been shown to induce MIPS\cite{Theurkauff2012}, there are other parameters that we expect would influence the resulting structure after a phase transition. The degree of particle softness has been shown to influence macroscopic properties of colloidal suspensions in equilibrium systems\cite{Vlassopoulos2012}, with investigations both \textit{via} theory\cite{Gnan2019TheColloids} and experiment\cite{Nigro2017,Seekell2015,Mattsson2009SoftGlasses}. Specifically, the influence of particle softness has been shown to alter the flow of the liquid phase\cite{Grand2008EffectsGlasses}, the conditions for glass formation (as well as its aging process\cite{Christopoulou2009AgeingGlasses}) and requires the reconsideration of the relevant driving forces (\textit{e.g.} the source of entropy) that determine phase behavior\cite{Vlassopoulos2012}. Moreover, experimentalists have demonstrated a great degree of control in synthesizing colloids of a specific softness, \textit{e.g.} by functionalizing colloids with polymers of different lengths and densities~\cite{Mahynski2015TuningMorphologies,Vlassopoulos2014TunableColloids}. Thus, a question arises, how does the rich behavior accessible by varying softness in Brownian colloidal systems transfer to active matter?

So far, no experimental studies have systematically investigated the effect of particle softness in active matter. Levis \textit{et al.} computationally examined the effect of particle softness and obtained a phase diagram relating activity and softness for four distinct repulsive strengths\cite{Levis2017}.
They found that making particles softer made the dense phase denser, and increased the threshold activity at which phase separation occurred. Additionally, various types of isotropic potentials have been examined: the Yukawa potential for soft particles\cite{Fily2012AthermalAlignment} or different strengths of the WCA potential\cite{Redner2016ClassicalAssembly,Stenhammar2014}) as well as anisotropic interactions \textit{e.g.} Janus interactions for Janus particles~\cite{Pu2017ReentrantInteraction}.

Most studies focus on different parameters that control the onset of MIPS. Here we focus, instead, on the structure of the dense phase and its interface with the gas phase after MIPS has occurred. The dense phase exhibits two spatial regions with distinct characteristics: a bulk dense phase and a dense-dilute interface \cite{Bialke2015a}. The bulk dense phase has constant density, whereas, the dense-dilute interface exhibits a monotonically decreasing density from the dense to the dilute phase density \cite{Bialke2015a}, resembling that of typical equilibrium liquid-gas interfaces \cite{Miyazaki1975, Weeks1977a, Chapela1977}. The stability of the dense phase is dictated by the balance of incoming and outgoing flux of particles from the cluster's surface. Incoming particles from the dilute phase are initially oriented towards the dense phase until rotational diffusion causes the particle's body axis to no longer be perpendicular to the cluster's surface \cite{Redner2013}. Alone, this would result in a rough interface lacking orientational alignment \cite{Barabasi1995}.  However, particles which bump into a rough interface will gradually move into convex regions of the surface, smoothing the interface and promoting local alignment \cite{Wysocki2014a, Fily2014, Nikola2016}.  This gives rise to a dense-dilute interface with a high degree of polarization of the body axes towards the cluster's center of mass, resulting in aligned body forces at the interface. To determine how these aligned body forces play a role in the mechanical stability of the steady state, we must first understand the momentum equation and its components.

\begin{table*}[t]
\large
\begin{tabularx}{\textwidth} { 
  | >{\centering\arraybackslash}b 
  | >{\centering\arraybackslash}s 
  | >{\centering\arraybackslash}s | }
\hline
\textbf{Simulation parameters} & \textbf{Definition} & \textbf{Value} \\
\hline
Particle diameter & $\sigma$ & $1.0$ \\ 
System size & $N$ & $10^5$ \\
System area fraction & $\phi=\frac{N\pi\sigma^2}{4 A_\text{box}}$ & [$0.45, 0.65$]\\ 
Interparticle interaction strength & $\epsilon$  & [$10^{-4}, 10^0$]\\ 
Rotational frequency & $\tau_\text{r}=D_\text{r}^{-1}$ & $\frac{1}{3}$ \\
Ratio of active to thermal forces (P\'eclet number) & $\mathrm{Pe}=\frac{3 v_\text{p}}{D_\text{r} \sigma}$ & [$0, 500$]\\ 
Ratio of active to pair potential forces & $F^\star=\frac{F^\text{a}\sigma}{ 24\epsilon}$ & [$10^0$, $10^6$] \\ 
\hline
\end{tabularx}
\caption{Definitions of important parameters in our analytical derivations in section \S2 and the values they take within our simulations in section \S3.}
\end{table*}

In Brownian suspensions and molecular liquids
the stress due to interparticle interactions is computed using virial formulae, which involves a volumetric integral of interparticle force moment \cite{Kirkwood1946, Irving1950}. 
This definition of stress recovers 
the Cauchy stress in continuum mechanics,
$\mathbf{\sigma}$, defined 
as a second-rank tensor that relates the 
traction vector, $\hat{\mathbf{F}}$ (force per unit area) on a surface 
with normal vector $\hat{\mathbf{n}}$ as
$\hat{\mathbf{F}}=\mathbf{\sigma}\cdot \hat{\mathbf{n}}$. Similarly the trace of the stress
tensor, defined as pressure, computed from 
determining the force per unit area of the surface 
and evaluating the volumetric integral yields the same results. 

Though the equivalence of interpreting physical processes from both a mechanical (microscopic) standpoint and a statistical mechanical (continuum-level) perspective applies for equilibrium systems, there has been ongoing debate 
about the appropriate microscopic formulation of stress in active suspensions, that is also consistent with a continuum definition. 
Brady and coworkers used the virial 
formulation to compute the average pressure
within a domain containing ABPs and showed that
the change in the direction of \emph{swimmers} due to interactions with the neighboring ABPs 
reduces the effective diffusivity of the swimmers and, thus, reduces the entropic stress. They referred to this activity-induced modification to pressure as \emph{swim pressure}, and used this quantity to predict the onset of MIPS in ABPs \cite{Takatori2015}. Consequent studies have shown that 
the pressure defined as the force per unit area on the boundaries of the computational domain is dependent on the detailed interactions of the particles with the boundary \cite{Solon2015a, Speck2016a}, leading to an argument that pressure is not a state variable in active systems. 

Surface tension, $\gamma$, similar to stress, is a
surface quantity and is defined as the energy required for creating a unit area of the interface \cite{Navascues1979}. Kirkwood and Buff \cite{Kirkwood1949} showed that, similar to stress, the surface tension in molecular liquids can be formulated as integrals
of interparticle forces over both phases 
and the interface. This formulation is consistent with the continuum  definition for equilibrium systems \cite{Brackbill1992}. 
Other studies have found that using a pressure 
formulation that contains swim pressure and
deploying Kirkwood and Buff formulation of surface tension results in extremely negative surface tension \cite{Bialke2015a, MariniBettoloMarconi2015, Paliwal2017, Patch2018, Solon2018b}. 

More recent studies \cite{Yan2015, Epstein2019, Omar2020MicroscopicMatter} have argued that these inconsistencies can be resolved if the swim pressure is not included in the stress calculations and instead the effect of particle activity in ABPs is modeled through a body force, due to 
the net alignment of ABPs at the interface, in the continuum limit. Particularly, Omar \textit{et al} showed that ignoring the swim pressure term 
leads to negligible surface tension in the dense-dilute interface of phase separated ABPs. 

In this paper, we analytically and computationally investigated the effect of softness for monodisperse active Brownian particles across a range of activities ($\mathrm{Pe}=3v_\text{p}\tau_\text{r}\mslash \sigma$ where $v_\text{p}$ is the intrinsic particle velocity, $\tau_\text{r}$ is the rotational frequency, and $\sigma$ is the particle diameter) and system area fractions, using the ABP model, see Table 1 for parameters details. We build upon the work of Levis \textit{et al.} \cite{Levis2017} by deriving analytical formulae that predict the resulting steady state structure of soft ABP systems. Focusing on the dense phase after MIPS has occurred, we describe two analytical approaches, a microscopic and a continuum one, built from few assumptions (average interaction between particles, hexagonal-close packing structure). We derived analytical expressions for the lattice spacing and the area fraction of both the bulk dense and dilute phase.  Then, in concert with kinetic theory, we obtained formulae for the cluster radius and the interparticle pressure. We found great agreement between analytical predictions and simulation results. To relate the microscopic pressure to the macroscopic pressure in the momentum equation, we explored the effect of particle softness, activity, and area fraction on surface tension. Consistent with the finding of refs.~\citenum{Epstein2019} and \citenum{Omar2020MicroscopicMatter}, we found that the swim pressure should not be included in the definition of the point-wise stress, and that the particle activity leads to a body force in the momentum equation near the interface. With these modifications, we derived a continuum approach for calculating the pressure arising from the aligned body forces at the interface (which approximately equals the interparticle pressure of the bulk dense phase), the interface width (which we find to be intrinsic to the system irrespective of varying particle softness, activity, or area fraction), and the surface tension. One important implication of our results is that across a range of parameters (softness, activity, system area fraction or size), the surface tension was found to be nearly zero and, therefore, play a negligible role in mechanically sustaining the steady state.

\begin{figure*}[t]
    \centering
    \includegraphics[width=0.95\textwidth,trim={0.05cm 0.2cm 0.08cm 0.0cm},clip]{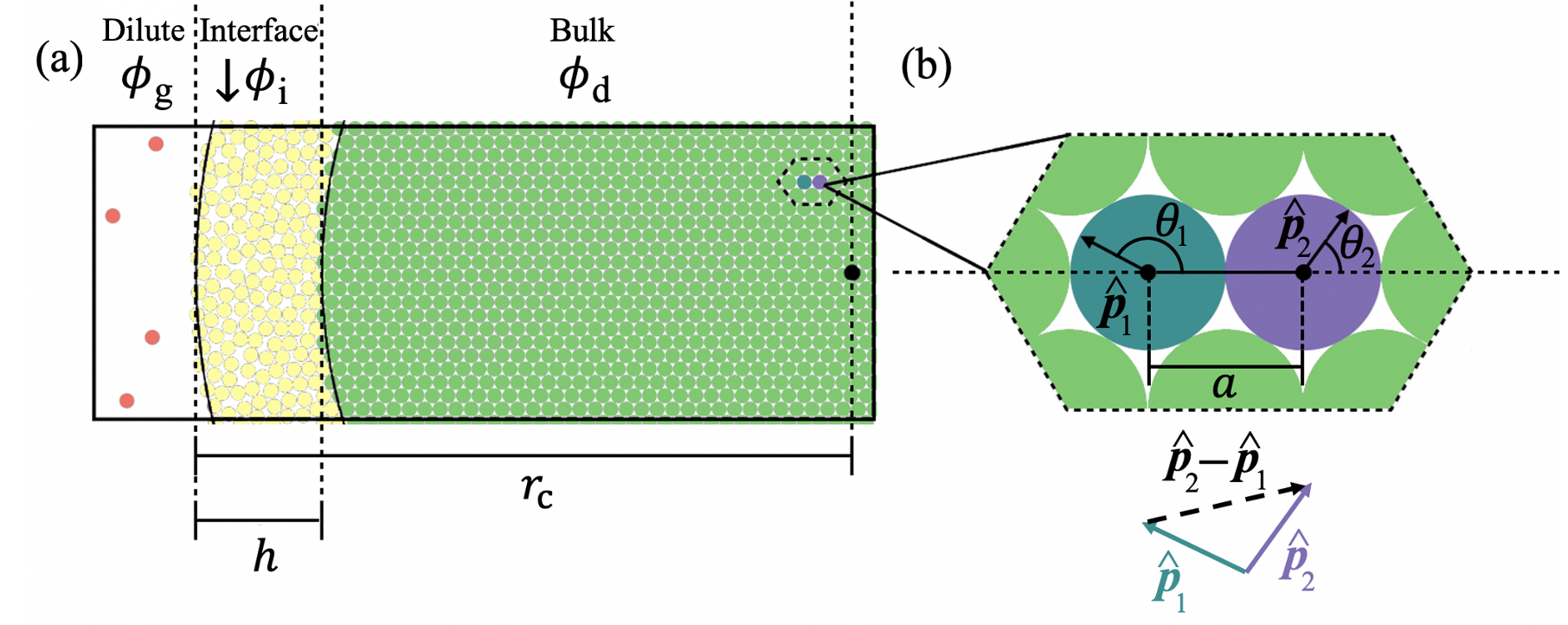}
    \caption{(a) Schematic representation of the cluster with three color-coded regions based upon the distance from the cluster's center of mass ($||\mathbf{r}||$): the bulk phase (green), the interface (yellow), and the dilute phase (red) with the cluster radius and interface width labeled ($r_\text{c}$ and $h$ respectively). The local area fraction of the bulk of the dense phase (for $0\le||\mathbf{r}||\le r_\text{c}-h$) is constant ($\phi_\text{d}$).  In the interface ($r_\text{c}-h \le ||\mathbf{r}|| \le r_\text{c}$), the local area fraction ($\phi_\text{i}$) decreases from the bulk phase area fraction, $\phi_\text{d}$, until it reaches the area fraction of the dilute phase outside the cluster, $\phi_\text{g}$. (b) Use of the pair force in computing the total force on a reference particle as a vector sum (eq.~\ref{eq:fpair_av}) over neighboring particles in a HCP structure, where $\hat{\mathbf{p}}_\text{i}$ indicates the $i^\text{th}$ particle's orientation, $\theta_\text{i}$ represents the angle between $\hat{\mathbf{p}}_\text{i}$ and the separation unit vector, $\hat{\mathbf{r}}$, and $a$ represents the average interparticle separation distance in the bulk of the dense phase, the lattice spacing.}
    \label{sch}
\end{figure*}

The structure of the paper is as follows. In section \S~\ref{analytical} we present our microscopic theoretical framework. In section \S~\ref{simulation}, we outline the simulation model and details for the systems studied here. We describe our results in section \S~\ref{results} showing comparisons between our analytical predictions and simulation results. In addition, we write down a continuum-theory approach and compare results with microscopic theory and simulations. Finally, we end with conclusions and outlook in section \S~\ref{conclusions}. 

\section{\S2: Theory} \label{analytical}

Consider a colloidal particle with a stiff repulsive core that is functionalized with a weakly repulsive polymer brush. In the extremely rarified case (a dilute gas), this colloid does not interact with neighbors and has a \lq resting\rq \ diameter, $\sigma$, when only thermal forces are present. However, upon increasing the system density, the functionalized colloid deforms due to an increasing number of interparticle interactions and has an effective diameter (less than $\sigma$) defined by the distance to its nearest neighbor, $||\mathbf{r}_\text{i}||$. With this kind of experimental system in mind we develop our analytical model of soft ABPs. 

Now, consider a system of ABPs which has undergone MIPS and is at steady-state: there is a dense phase and a dilute (gas-like) phase (fig.~\ref{sch}a).  In what follows, we will be focusing on the dense phase and will be calculating the lattice spacing, area fraction, cluster radius, interparticle pressure, and the surface tension based on a small number of assumptions, discussed first. 

In the dense phase, the velocity of particles is 
negligible compared to the velocity of particles in the dilute phase. Thus, we assume that the particle hydrodynamic drag forces (which are proportional to the particle velocity) are also negligible. In accordance with the previous and our own findings from simulations, we also assume that the particles in the dense phase are arranged in a hexagonally close-packed (HCP) lattice \cite{Redner2013} (see fig.~\ref{sch}b and Electronic Supplementary Information (SI), fig. S1). The lattice spacing ($a$) between neighboring particles will be determined through a balance of the neighbors' active forces (which compress a particle) and the repulsive forces (which resist particle overlap). First, we distinguish two regions within the dense phase: the bulk, which includes the majority of particles in the dense phase, and whose constituent particles' body axes exhibit no orientational alignment on average, and the interface between the dense and dilute phases, where particles possess a high degree of orientational alignment towards the cluster's center of mass. Note that the existence of a bulk dense phase and a dense-dilute interface is supported by previous works \cite{Paliwal2018, Solon2018, Hermann2019} as well as our own simulation results presented in section \S~\ref{results} fig.~\ref{radial}. As such, the compression of particles within the bulk results from the aligned particles at the edge of the dense phase, \textit{i.e.} the interface pushing inward towards the center of the cluster. As a result, the bulk particles' effective diameter is smaller than the resting diameter. The effective diameter, which is equal to the interparticle separation of immediate neighbors, has little variability within the bulk dense phase.  Therefore, we assume that each particle within the bulk dense phase is a constant distance apart from its nearest neighbors, equal to the lattice spacing, $a$. 

Based on these assumptions, our first aim is to analytically compute the different structural and mechanical parameters of the dense phase, including its lattice spacing, cluster radius and interparticle pressure, at a variety of activities and stiffnesses (repulsive strengths).
Our particles interact through the Weeks-Chandler-Andersen (WCA) potential

\begin{equation}\label{ljPotential}
U({r_\text{i,j}}) = 
	\begin{cases} 
      4\epsilon [(\frac{\sigma}{{r_\text{i,j}}})^{12} - (\frac{\sigma}{{r_\text{i,j}}})^6]+\epsilon & 0 \leq {r_\text{i,j}}\leq \sqrt[\leftroot{-3}\uproot{3}6]{2}\sigma \\
      0 & {r_\text{i,j}}> \sqrt[\leftroot{-3}\uproot{3}6]{2}\sigma, \\
   \end{cases}
\end{equation}
which provides repulsion at distances up to slightly greater than the resting particle diameter and is zero beyond that distance $(r_\text{cut}=\sqrt[\leftroot{-3}\uproot{3}6]{2}\sigma)$. 
Here $\sigma$ defines the resting particle diameter when only thermal forces are present, $\epsilon$ determines the interaction strength, and $r_\text{i,j} = ||{\mathbf{r}}_\text{j} - {\mathbf{r}}_\text{i}||$ is the center-to-center separation between two particles $\mathrm{i}$ and $\mathrm{j}$. 
The interparticle force applied by particle $\mathrm{j}$ on particle $\mathrm{i}$ is the gradient of this potential, $\mathbf{F}^{\text{WCA}}(\mathbf{r})=-\nabla_\mathbf{r} U$, and is given by:

\begin{equation}\label{ljForce}
\mathbf{F}^{\text{WCA}}(r_\text{i,j}) =
	\begin{cases} 
      \frac{24\epsilon}{\sigma} [2(\frac{\sigma}{r_\text{i,j}})^{13} - (\frac{\sigma}{r_\text{i,j}})^7] \hat{\mathbf{r}}& 0 \leq {r_\text{i,j}}\leq \sqrt[\leftroot{-3}\uproot{3}6]{2}\sigma \\
      0 & {r_\text{i,j}}> \sqrt[\leftroot{-3}\uproot{3}6]{2}\sigma, \\
   \end{cases}
\end{equation}
where $\mathbf{r}_\text{i,j}=\mathbf{r}_\text{j}-\mathbf{r}_\text{i}$ and $\hat{\mathbf{r}}=\mathbf{r}_\text{i,j}/||\mathbf{r}_\text{i,j}||$ is the relative separation unit vector. Note that particle stiffness is modulated via the interaction strength, $\epsilon$, where larger $\epsilon$ corresponds to stronger repulsive forces and, in turn, stiffer particles.

Let us begin by computing the active force exerted by an isolated pair within the dense phase; see fig.~\ref{sch}b. The average force applied by particle 2 on particle 1 can be generally expressed as 
\begin{equation}
     \langle\mathbf{F}^\text{pair}\rangle= \int \mathbf{F}^\text{pair} P(\hat{\mathbf{p}}_1, \hat{\mathbf{p}}_2)d\hat{\mathbf{p}}_1 d\hat{\mathbf{p}}_2,
    \label{eq:fpair_av}
\end{equation}
where $P(\hat{\mathbf{p}}_1, \hat{\mathbf{p}}_2)$ is the probability density function of observing particles 1 and 2 at orientations $\hat{\mathbf{p}}_1$ and $\hat{\mathbf{p}}_2$, respectively; see fig.~\ref{sch}b. We know that in the bulk dense phase the orientations of the particles are independent of each other and are uniformly distributed:  
$P(\hat{\mathbf{p}}_1, \hat{\mathbf{p}}_2)=P(\hat{\mathbf{p}}_1)P(\hat{\mathbf{p}}_2)=(1/2\pi)^2$. The pair force is nonzero only when the relative motion of the pair causes overlap. Therefore, since the large active force ($\mathbf{F}^\text{a}$) dominates over translational Brownian fluctuations, the pair force, $\mathbf{F}^\text{pair}$, is only a function of activity and interparticle forces. In the absence of any orientational anisotropy,  the pair force acts only along the line of centers of the particles, $\hat{\mathbf{r}}$, and is equal to the projection of the pair relative velocity in the $\hat{\mathbf{r}}$ direction:
\begin{equation}
\mathbf{F}^\text{pair}_1=F^\text{a}
    \begin{cases}
    \left[\hat{\mathbf{r}}\cdot \left(\hat{\mathbf{p}}_2-\hat{\mathbf{p}}_1\right)\right]\hat{\mathbf{r}} & \hat{\mathbf{r}}\cdot \left(\hat{\mathbf{p}}_2-\hat{\mathbf{p}}_1\right)<0, \\
    \mathbf{0} & \hat{\mathbf{r}}\cdot \left(\hat{\mathbf{p}}_2-\hat{\mathbf{p}}_1\right)\ge 0 
    \end{cases}
    \label{eq:fpair}
\end{equation}

\noindent Substituting $P(\hat{\mathbf{p}}_1, \hat{\mathbf{p}}_2)={(1/2\pi)}^2$ and eq.~\ref{eq:fpair} into eq.~\ref{eq:fpair_av}, we simplify to find the average pair force experienced by an isolated pair of particles in a dilute system
\begin{equation} \label{eq:dilutepredict}
     \langle\mathbf{F}^\text{pair}\rangle = 4 F^\text{a} \hat{\mathbf{r}}{(\frac{1}{2\pi})}^2 \int_0^\pi \int_0^{\theta_1} \left(\cos \theta_2-\cos \theta_1\right)d\theta_2d\theta_1=\frac{4}{\pi^2} F^\text{a} \hat{\mathbf{r}}. 
\end{equation}

\noindent This calculation, however, ignores the effect of surrounding ``bath'' particles on $\langle\mathbf{F}^\text{pair}\rangle$, which may dominate the pair interactions at large area fractions. Thus, instead of using the prefactor $4/\pi^2$, 
we assume the general form  $\langle\mathbf{F}^\text{pair}(a)\rangle=\beta {F}^\text{a}\hat{\mathbf{r}}$. 

\begin{figure}
\centering
  \includegraphics[width=0.45\textwidth,trim={1.2cm 1.2cm 1.3cm 1.4cm},clip]{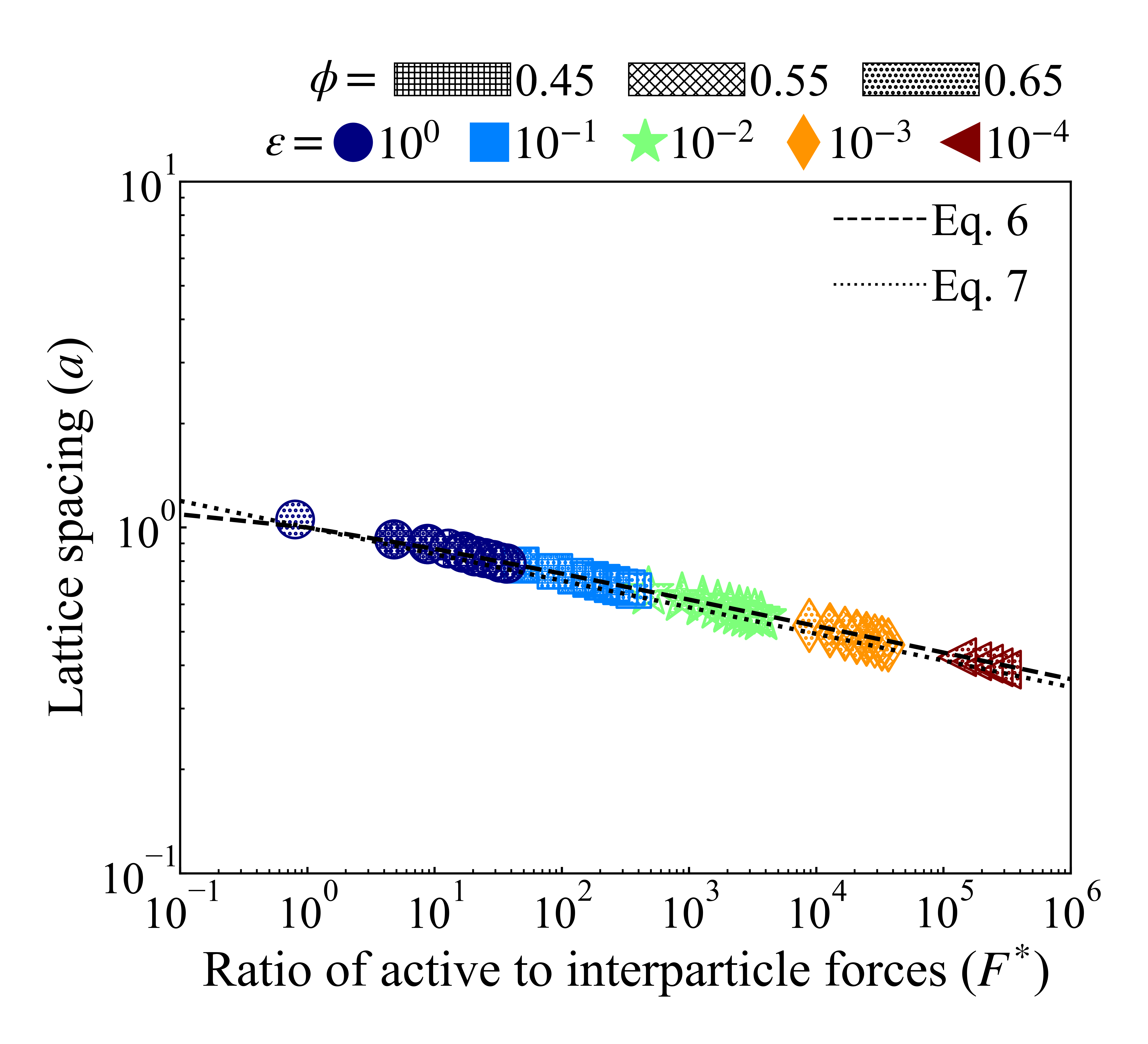}
  \caption{Lattice spacing ($a$) of the HCP phase for variably soft ABPs at distinct, constant, potential well depth ($\epsilon$, evenly spaced on log-scale, see legend) for both simulation (points of varying shape and color, see legend) and with the best fit using $\beta=2.0$ (discussed in detail in the SI, see section \S 1) with eq.~\ref{eq:fstar} (dashed lines) and eq.~\ref{eq:fstar_red} (dotted lines). Increasing activity ($\mathrm{Pe}\propto F^*$) or softness (decreasing $\epsilon$) corresponds to a shorter lattice spacing. Varying system area fraction (different hatching, see legend) has negligible effect on the lattice spacing at constant softness and activity. Furthermore, all data collapses to a single curve in log-log scale. We find the simplified eq.~\ref{eq:fstar_red} with $\beta=2.0$ reliably agrees with that derived analytically in eq.~\ref{eq:fstar} and measured via simulation.}
  \label{lat_derive}
\end{figure}

Since the dense phase particles are assumed to be static, the pair active force ($\langle\mathbf{F}^{\text{pair}}\rangle$) and the repulsive interparticle force ($\mathbf{F}^\text{WCA}$) must be equal, giving a force balance equation that enables the determination of the lattice spacing in the dense phase ($a$): 

\begin{equation}
    \beta F^\star = 2\left(\frac{\sigma}{a}\right)^{13} -\left(\frac{\sigma}{a}\right)^7,
    \label{eq:fstar}
\end{equation}
where $F^\star=\frac{F^\text{a} \sigma}{24\epsilon}$ is the ratio of active to interparticle forces. We, then, proceed to use simulation results to compute $\beta$. Fig.~\ref{lat_derive} shows the simulation values of $a$ vs $F^\star$ using eq.~\ref{eq:fstar} for a wide range of $\mathrm{Pe}$ and $\epsilon$. 
The data collapse into a single curve. The dashed lines shows the fit from eq.~\ref{eq:fstar} for $\beta=2.0$, which is in excellent agreement with simulation results (discussed in detail in the SI, see section \S 1). To simply things further, we neglect the second term on the RHS of eq.~\ref{eq:fstar} and compute $\sigma/a$ as

\begin{equation}
    \frac{\sigma}{a} \approx \left(\frac{\beta F^\star}{2}\right)^{1/13}.
    \label{eq:fstar_red}
\end{equation}.

We find the simplified eq.~\ref{eq:fstar_red}, plotted as a dotted line using $\beta=2.0$ in Fig.~\ref{lat_derive}, is in almost equally strong agreement with our simulation data as eq.~\ref{eq:fstar}. Note that $\beta=2$ is approximately $5$ times larger than 
the computed value, when the effect of bath particles is neglected ($4/\pi^2$), indicating the dominant role of bath particles in determining the effective pair interactions; this is to be expected in this range of area fractions (discussed in detail in the SI, see section \S 1). 

Knowing the lattice spacing $a$ enables us to determine the area fraction of both the dense and dilute phases, which allow for the calculation of three important quantities: i) a binodal of the dense and dilute phase area fractions, ii) the number of particles in the dense phase and iii) the radius of the cluster at steady-state. The dense phase area fraction can be calculated from: 

\begin{equation}\label{phid}
\phi_\text{d}=\frac{\phi_\text{cp}\sigma^{2}}{a^{2}},
\end{equation}

\noindent where $\phi_\text{cp}=\pi/(2\sqrt{3})$ is the area fraction of disks in a HCP lattice. Note that because our particles are soft they can compress so that the lattice spacing ($a$) is smaller than the particle diameter (Fig.~\ref{lat_derive}) and the dense phase area fraction is greater than close packing, $\phi_\text{cp}$. In order to obtain the area fraction of the dilute phase we follow Redner \textit{et al.}\cite{Redner2013} and define rates of adsorption on to ($k_\text{in}$) and desorption from ($k_\text{out}$) the dense-phase cluster:
\doubleequation[kin,kout]{k_\text{in} = \frac{n_\text{g}v_\text{p}}{\pi}}{k_\text{out} = \frac{\kappa D_\text{r}}{a},} \label{doubleeq}
\noindent where $n_\text{g}$ is the number density of the dilute phase, $\nu_\text{p}$ is the swim velocity of a single particle, $D_\text{r}$ is the rotational diffusion rate, and $a$ is the lattice spacing of the dense phase,  as predicted by eq.~(\ref{eq:fstar}). 
Redner \textit{et al.}\cite{Redner2013} used the fitting parameter $\kappa=4.5$ explaining it stems from the observation that particle desorption occurs in avalanche-like events. At steady state, the rate of adsorption is equal to the rate of desorption,
$k_\text{in}=k_\text{out}$, which gives

\begin{equation}\label{equalRate}
    n_\text{g} = \frac{\pi \kappa D_\text{r}}{\nu_\text{p}a}.
\end{equation}

\noindent Multiplying eq.~\ref{equalRate} by the area of a particle ($A_\text{p}=\pi\sigma^2$), we can present this quantity in terms of typical input parameters, namely the area fraction and activity,

\begin{equation}\label{phiG}
    \phi_\text{g} = \left(\frac{3\pi^{2}\kappa}{4}\right)\left(\frac{a}{\sigma} \mathrm{Pe}\right)^{-1}, 
\end{equation}
where $\mathrm{Pe}=\frac{3 v_\text{p}}{D_\text{r} \sigma}$ is the ratio of active to thermal forces. 

So far, we have calculated the area fraction of both dense (eq.~\ref{phid}) and dilute (eq.~\ref{phiG}) phases. We can also compute the number of particles in the dense phase in terms of $\phi$, $\phi_\text{g}$, and $\phi_\text{d}$ given the system size ($N$), simulation box area ($A$), and the system area fraction ($\phi=NA_\text{p}/A$), which are all known inputs for our simulations.  Further simplification using eqs.~\ref{phid} and \ref{phiG} enables us to calculate $N_\text{d}$ (Lever rule) based upon our physical input parameters of $\phi$, $\mathrm{Pe}$, and $\epsilon$, (discussed in detail in the SI, see section \S2),

\begin{align}
    N_\text{d} &= N\left( \frac{\phi_\text{d}(\phi_\text{g}-\phi )}{\phi (\phi_\text{g}-\phi_\text{d})} \right) \nonumber \\
    &= N \left( \frac{\phi_\text{cp}\sigma^{2}}{\phi a^{2}} \right)
    \frac{ \phi-\frac{3\pi^{2}\kappa\sigma}{4}\left( \frac{a}{\sigma} \mathrm{Pe}\right)^{-1}}
    {\frac{\phi_\text{cp}\sigma^{2}}{a^{2}}-\frac{3\pi^{2}\kappa\sigma}{4}\left( \frac{a}{\sigma} \mathrm{Pe}\right)^{-1}} \quad . \label{nl}
\end{align}

The area of the dense phase cluster, $A_\text{d}$  can be expressed as a function of the effective particle diameter ($a$), the number of dense-phase particles ($N_\text{d}$), and the packing fraction of the HCP lattice ($\phi_\text{cp}$):
\begin{equation}\label{ac}
    A_\text{d}=\frac{N_\text{d}\pi a^{2}}{4 \phi_\text{cp}}.
\end{equation}

\noindent Next, we compute the cluster radius ($r_\text{c}$) as a function of activity ($\mathrm{Pe}$), softness (\textit{via} $a$), system area fraction ($\phi$), and resting particle diameter ($\sigma$):

\begin{equation}\label{rc}
    r_\text{c}=\sqrt{\frac{N_\text{d}}{4 \phi_\text{cp}}}a.
\end{equation}

We now seek to compute the interparticle pressure within the dense phase using the virial formulation,

\begin{equation}
    \Pi^\text{P}_\text{d}=n_\text{d} \hat{\Pi}^\text{p}_\text{d}=n_\text{d}\frac{1}{2}\sum^{N=6}_{i=1} \mathbf{r}_\text{i} \cdot \mathbf{F}^\text{WCA}(a),
    \label{eq:pressure1}
\end{equation}
where $\mathbf{r}_\text{i}$ is the position vector between the centrally-tagged reference particle and its $\text{i}^{th}$ neighbor, $n_\text{d}=\frac{N_\text{d}}{A_\text{c}}$ is the number density of the dense phase, $\hat{\Pi}^\text{P}_\text{d}$ is the interparticle pressure on a single particle in the dense phase, the superscript $\text{P}$ denotes interparticle interactions. 
Using eq.~\ref{ac} and $F^\text{WCA}(a)=2F^\text{a}$ simplifies eq.~(\ref{eq:pressure1}) to 
\begin{equation}
    \Pi^\text{P}_\text{d}=\frac{2\beta\sqrt{3}F^\text{a}}{a}. 
    \label{eq:pressureVP}
\end{equation}

\noindent The form of pressure in eq.~\ref{eq:pressureVP} is not immediately intuitive. The difficulty arises since we are studying the system after MIPS, where the particles are forming a crystalline phase. Our simulations and theory show that the pressure (and other variables) of the crystalline phase is independent of the initial area fraction for $\phi\ge0.45$, even though $\phi$ is a determinant of the onset of MIPS. Below this area fraction, we did not observe the transition to the crystalline phase. Thus, the dimensionless pressure in our system is only a function of activity ($F^\text{a}$) and interparticle interactions ($\epsilon$, $\sigma$). To make eq.~\ref{eq:pressureVP} more intuitive, we now explicitly present the pressure in its dimensionless form. To do so, we substitute eq.~\ref{eq:fstar_red} in for $a$ in eq.~\ref{eq:pressureVP},

\begin{equation}
    \frac{\Pi^\text{P}_\text{d}}{\Pi_0}=2\beta\sqrt{3} \left(\frac{\beta F^\star}{2}\right)^{1/13}
\end{equation},

\noindent where $\Pi_0=F^\text{a}/\sigma$. This equation is only a function of $F^\text{a}$, $\epsilon$, $\sigma$ and independent of $k_\text{B} T$ and $\phi$.

Recall that we are interested in the limit of $\mathrm{Pe} \gg 1$, beyond the MIPS critical point. Considering this  
limit in eq.~\ref{phiG} and \ref{nl} gives $\phi_\text{g}\to 0$ and $N_\text{d}\to N$ \textit{i.e.} all particles will be adsorbed to the dense phase. Similarly, taking $\mathrm{Pe}\gg 1$ and writing eq.\ref{ac} in terms of $r_\text{c}$, we find 
that the radius of the dense phase scales linearly with the lattice spacing and the dimensions of the simulation box ($L_\text{box}\propto \sqrt{N}$):
\begin{equation}\label{newRad}
    \frac{r_\text{c}}{a}=\left(\sqrt{\frac{\sqrt{3}}{2\pi}}\right)\sqrt{N}.
\end{equation}

To sum up, we have given analytical expressions for 
the macroscopic mechanical variables, including interparticle pressure in the dense phase, as well as  microstructural variables, including the lattice spacing, $a$, and the radius of the dense phase, $r_\text{c}$, in terms of activity $\mathrm{Pe}$, softness $\epsilon$, resting particle diameter $\sigma$ and area fraction $\phi$. Our only assumptions were that the dense phase forms an HCP lattice, that the activity is high and dominates over Brownian motion, and the only interparticle forces we consider are from immediate neighbors. To test the validity of our analytical calculation, next we compare our predictions against the results from Brownian Dynamics simulations. 

\section{\S3: Simulation Methods} \label{simulation}

\begin{figure}
   \includegraphics[width=0.45\textwidth,trim={0 0 0 0},clip]{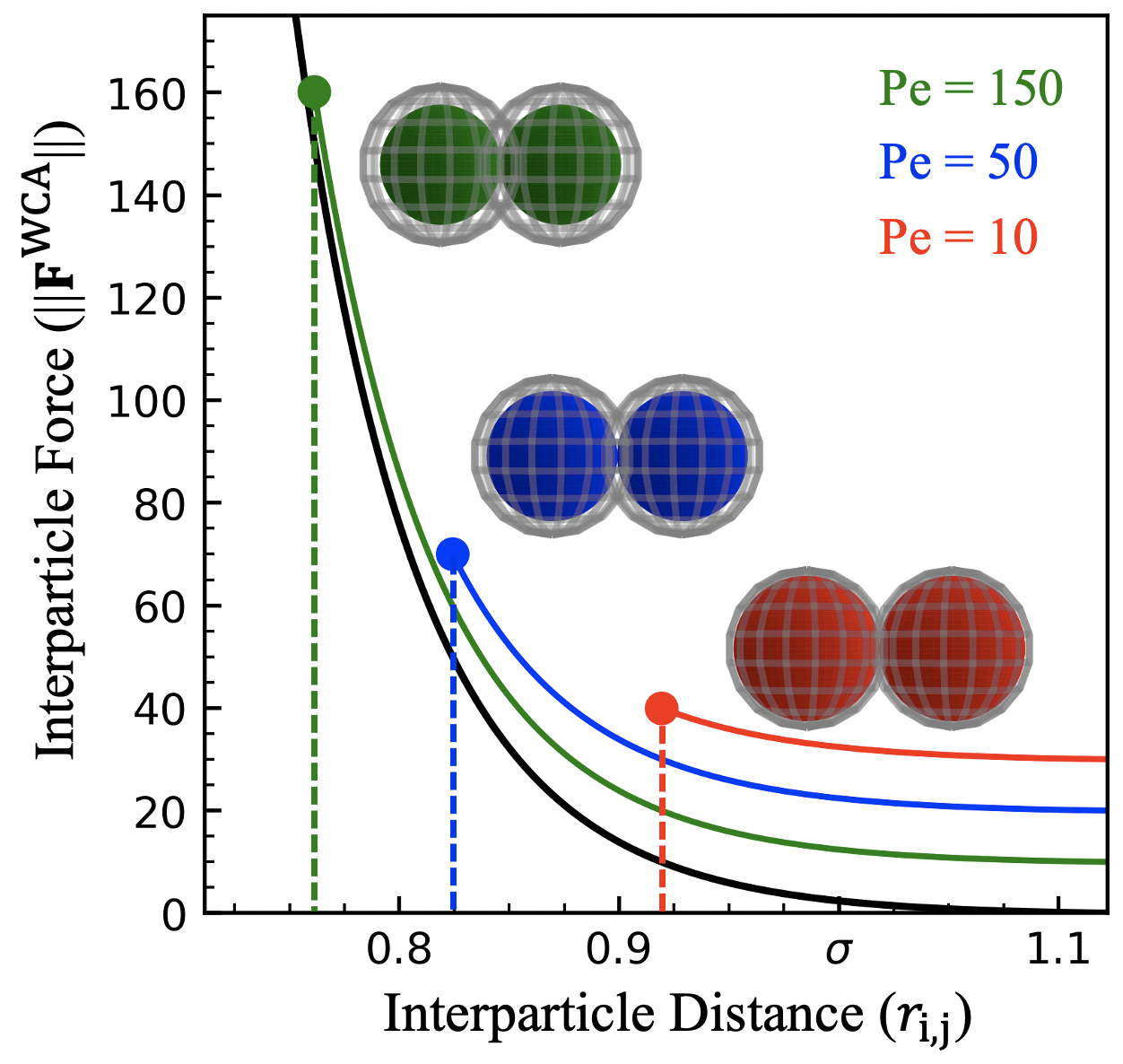}
  \caption{Force at varying interparticle separation distance ($r_\text{i,j}$ for the WCA potential (eq.~\ref{ljPotential}, black) for constant repulsive well-depth $(\epsilon=0.1)$, showing sensitivity to activity. For a head-on collision of active particles
($\mathrm{Pe}_\text{red}<\mathrm{Pe}_\text{blue}<\mathrm{Pe}_\text{green}$) points indicate the maximum particle deformation while lines indicate range of interparticle distance available in other types of collisions. Colored spheres demonstrate the maximum deformation at each activity and wire mesh overlay shows the extent of particle overlap (for particle diameter $\sigma=1.0$).}
  \label{schematic}
\end{figure}

We simulate $N=10^5$ spherical active Brownian particles (ABPs), of diameter $\sigma$, confined to a  two-dimensional simulation box $L_\text{box}\propto\sqrt{N}$ and subject to periodic boundary conditions. Each particle's activity is modulated by varying the active force (${\mathbf{F}}^\text{a}=\xi v_\text{p}\widehat{\mathbf{p}}$ with drag $\xi$ from an implicit solvent), which is applied along a body axis, or swim direction, defined through unit vector $\widehat{\mathbf{p}}=(\cos{\varphi},\sin{\varphi})$, where $\varphi$ is the angle between the body axis and the positive $x$-axis. Note that the drag is dependent on the effective particle area which varies with a particle's effective diameter and, in turn, softness. Despite changing the kinetics, variations in drag (and, in turn, the rotational and translational diffusion coefficients indirectly) would negligibly influence our predictions as the structure of the dense phase is static and, hence, independent of hydrodynamic interactions. The active force is varied \textit{via} the intrinsic particle velocity ($v_\mathrm{p}$). A particle's motility, or activity, is quantified by the (dimensionless) P\'eclet number $(\mathrm{Pe}=3v_\text{p}\tau_\text{r}\mslash \sigma)$. We ensure our results are not influenced by finite-size effects from the periodic boundary conditions by testing different system sizes, see SI, figs.~S5,S9. 

Particles translate and rotate in accordance with Brownian dynamics,

\begin{equation}\label{brownmotion}
\dot{\mathbf{r}}_\text{i}= \frac{1}{\xi}\left(\mathbf{F}_\text{i}^\text{WCA} +F^\text{a}\widehat{\mathbf{p}_\text{i}}\right)+ \sqrt{2D_\text{t}}\bm{\Lambda}_\text{i}
\end{equation}
\begin{equation}
\dot{\varphi_\text{i}}=\sqrt{2D_\text{r}}\,\Gamma_\text{i} \quad,
\end{equation}

\noindent where, $\mathbf{r}_\text{i}$ provides the $\text{i}^{th}$ particle's position in two dimensions, ${\bm{\Lambda}}_\text{i}$, and $\Gamma_\text{i}$ represent zero-mean unit variance Gaussian noise, $\left\langle {\Lambda}_\text{i}^{\alpha}(t)  {\Lambda}_\text{j}^{\beta}(t')  \right\rangle = {\delta}_\text{ij} {\delta}_{\alpha \beta} \delta (t-t')$, $\left\langle {\Gamma}_\text{i}(t)  {\Gamma}_\text{j} (t')  \right\rangle = {\delta}_\text{ij} \delta (t-t')$, where $\alpha$ and $\beta$ denote Cartesian coordinates, and ${\mathbf{F}_\text{i}}^\text{WCA} = -\sum_{\text{j}\neq \text{i}}{\nabla}_\text{i} U({r}_\text{ij})$ is the repulsive interparticle force from the WCA potential (eq.~\ref{ljPotential}). Drag $(\xi=3\pi\eta\sigma)$ and translational/rotational noise ($D_\text{t}=k_\text{B}T\mslash \xi$, $D_\text{r}=3D_\text{t}\mslash \sigma^{2}$) are set according to system temperature $(T)$ and particle size (resting particle diameter $\sigma$) for a given fluid with dynamic viscosity $(\eta)$. In addition to sudden orientation changes from collisions, the body-axis reorients randomly over time according to a characteristic timescale $\tau_\text{r}=D_\text{r}^{-1}=1/3$. Note that our implementation and results apply also to systems where random translational and rotational motion do not stem from the system temperature. That is to say, the emergent phenomena we observed result from the relationship between a particle's velocity and rotational frequency ($\tau_\text{r}=D_\text{r}^{-1}$), which is encoded in the persistence length ($l_\text{p}=v_\text{p}\tau_\text{r}$). Thus, the rotational frequency need not be explicitly thermal in origin (and could be set to any reorientation rate that reflects a particular system).

Interaction between particles is described through a WCA potential (eq.~\ref{ljPotential}). We implement several values of the repulsive well depth $(\epsilon)$ to study the effect of softness on the structure and mechanics of the dense phase. Recall that at a fixed interparticle distance, a larger value of $\epsilon$ produces a greater repulsive force \textit{i.e.} hard particles have a larger value of the repulsive well depth than soft particles $(\epsilon_{\text{hard}}>\epsilon_{\text{soft}})$. For a constant repulsive strength $(\epsilon)$, increasing the activity (and therefore the active force) results in greater particle overlap (fig.~\ref{schematic}). Despite using a constant value of softness for particles in a given simulation, there is a distribution of effective particle diameters resulting from a different degree of particle compression. Depending on the environment of any given particle, the degree of compression will vary according to its neighbors' orientations and the resulting compressive forces acting on it. An experimental analogue to this implementation of the interparticle potential could be a colloid functionalized with a polymer brush where distinct repulsive strengths can be viewed as a brush of different length and density~\cite{Vlassopoulos2012}.

 We used the molecular dynamics package HOOMD-Blue\cite{Anderson2020, Glaser2015, Anderson2008} to simulate $N=10^{5}$ monodisperse active particles for a simulation time interval of $\tau=300\tau_\text{r}$ ensuring that steady state had been reached. We varied: system area fraction $(\phi=0.45, 0.55, 0.65)$, particle activity $(\mathrm{Pe}=50-500$ in steps of $50$), and the potential well depth resulting in different softness ($\epsilon=1, 10^{-1}, 10^{-2}, 10^{-3}, 10^{-4} k_\text{B}T$). We focus on systems that phase separate via MIPS into dense and dilute phases (see fig.~\ref{phase}). To overcome kinetic limitations of cluster formation, we instantiate small circular clusters (discussed in detail in the SI, see section \S 3). We note that the steady-state composition of a cluster is independent of its initial seeded size (see SI, fig. S1 and S2). As in the analytical approach, the steady-state dense phase is comprised of a bulk domain in the interior and an interface that separates the bulk dense phase from the gas phase. 
 
\section{\S4: Results} \label{results}

\subsection{\S4.1: Properties of the dense phase: gas, bulk, interface}\label{clustProps}

\begin{figure}
\centering
  \includegraphics[width=0.45\textwidth, trim={0.2cm 2.7cm 2.3cm 2.0cm},clip]{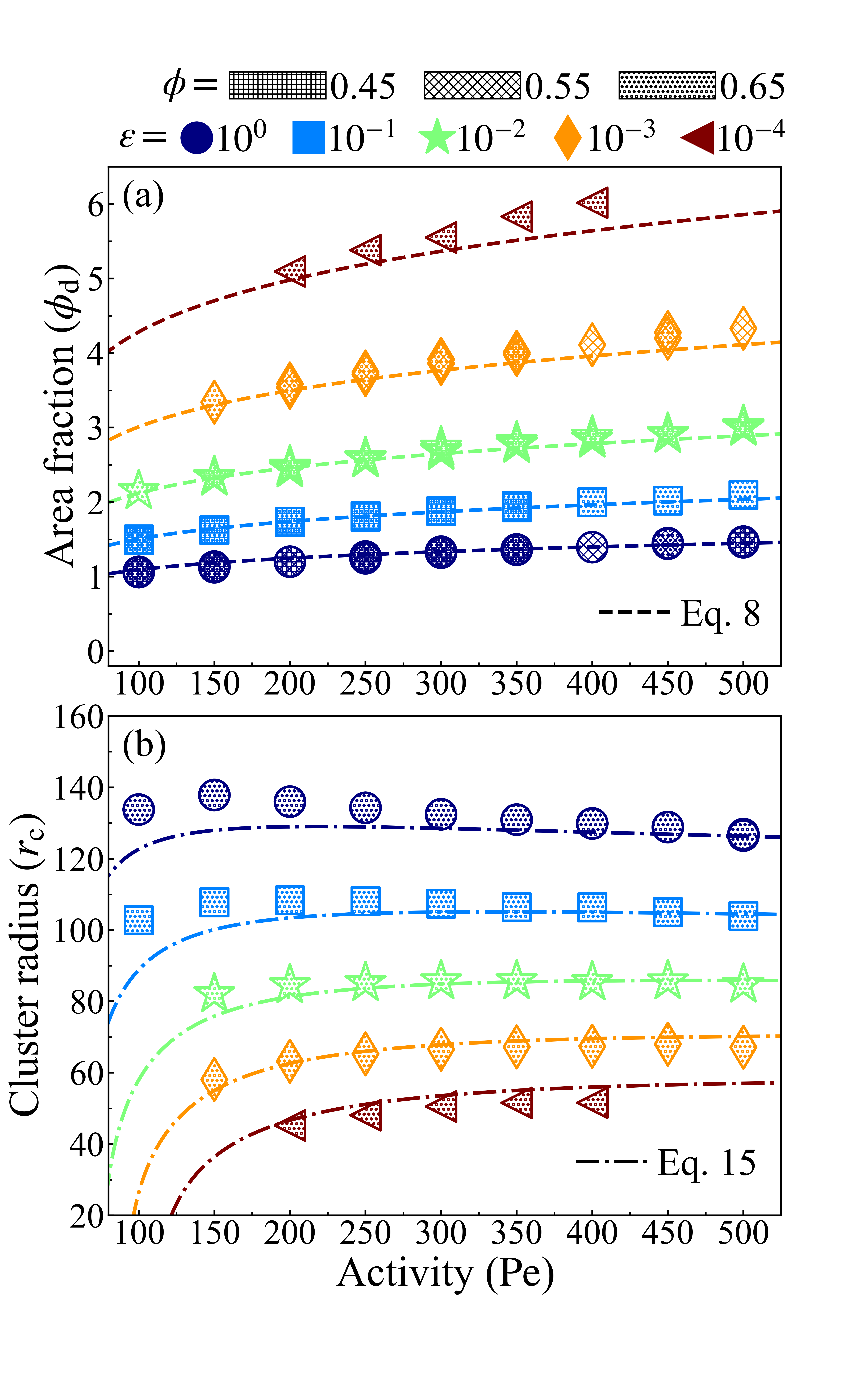}
  \caption{(a) Corresponding local area fraction of the dense phase from simulation data is computed as a local bin and shows sound agreement with analytical approach. The dense phase becomes more densely packed (and, in turn, the gas phase becomes more dilute) via increasing particle softness or activity (at constant softness). (b) Analytically computed cluster radius from eq.~\ref{rc} at system area fraction of $\phi=0.65$ with simulations at various softness (color). Strong agreement between the simulation values and the analytical results are seen for all other system area fractions tested ($\phi=0.45$ and $\phi=0.55$).}
  \label{phase}
\end{figure}

\begin{figure*}[bt!]
\centering
  \includegraphics[width=1.0\textwidth]{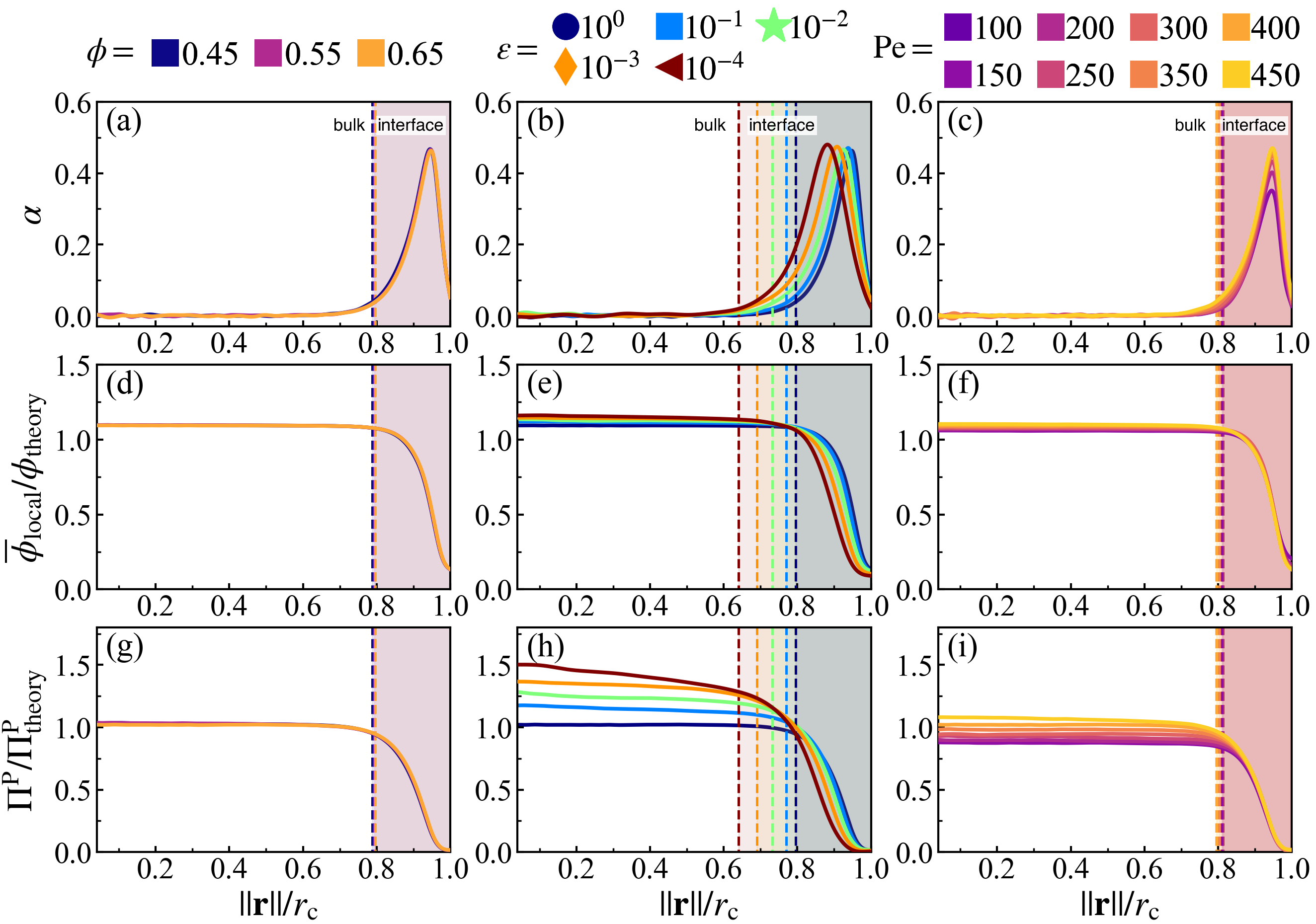}
  \caption{Orientational alignment towards the cluster's center of mass ($\alpha(r)=\overline{-\hat{\mathbf{p}}(r)\cdot \mathbf{r}}>0$) (a-c), local area fraction (d-f), and interparticle pressure (g-i) calculated using the virial formulation of pressure (eq.~\ref{eq:pressure1}).  Data is radially binned and measured over twenty $18^\circ$ conical surfaces per time step. As evident in the x-axis, the radial location ($||\mathbf{r}||$) is normalized by the measured cluster radius ($r_\text{c}$) of each conical surface, and all data is averaged over time at steady-state (for at least $50\tau_\text{r}$) and all conical surfaces (twenty conical surfaces per time step). Similarly, the measured local area fraction and pressure are normalized by dividing through by their analytically predicted values (eqs. \ref{phid} and \ref{eq:pressureVP}), respectively. Data shows the effect of both system area fraction (a,d,g, see legend), softness (b,e,h, see legend), and activity (c.f.i, see legend). For each column, the parameter being varied is in the above legend while the other two system parameters are held constant ($\mathrm{Pe}=350$, $\epsilon=1.0$, or $\phi=0.65$).  }
  \label{radial}
\end{figure*}

The dense phase cluster is highly dynamic, i.e. it frequently changes size (see SI, fig. S3-6) and shape (see SI, fig. S7-10), with some parts breaking up into smaller clusters and merging back, similar to references~[\citenum{Redner2013, Theurkauff2012}]. As the activity is increased the fluctuations of the interface are decreased leading to a more stable shape (see SI, fig. S7). In the analysis that follows, we are concerned with the dynamics of the largest length-scale/wavelengths (i.e. cluster radius) and sufficiently large activities that lead to crystallization of the dense phase. Specifically, we are not concerned with interface fluctuations at shorter wavelengths (the amplitude of these fluctuations is less than 1\% of the cluster radius, see SI, fig. S7-10). 

How do the properties of the dense phase, such as cluster size and pressure, change when the particles become softer? Levis \textit{et al.} computationally calculated the phase diagram for ABPs at various particle softnesses and found a softness-dependent binodal~\cite{Levis2017}. They showed that soft particles undergo MIPS at a smaller critical cluster size than hard particles; however, due to a lower nucleation barrier, softer clusters could more easily destabilize and break apart, necessitating larger activities or area fractions for sustained phase separation~\cite{Levis2017}. Here, we explore a broader range of repulsive strengths ($\epsilon$), activities ($\mathrm{Pe}$) and system area fractions ($\phi$) and compare both with our analytical predictions and the observations of Levis \textit{et al.} \cite{Levis2017}. 

Our simulation results, in agreement with Levis et al~\cite{Levis2017}, show that softer particles pack more densely and therefore shift the dense phase of the binodal to higher densities at constant activity, (fig.~\ref{lat_derive}). At fixed softness, increasing particle activity also makes the dense phase denser and reduces the lattice spacing (fig.~\ref{lat_derive}). Increasing the system area fraction ($\phi$) at values greater than the critical area fraction has negligible influence on the area fraction of the dense phase, as we see nearly perfect overlap of points with constant activity ($\mathrm{Pe}$) and softness ($\epsilon$) (fig.~\ref{phase}a). Note that the theoretical predictions of $\phi_\text{d}$ from eq.~\ref{phid} are in good agreement with the simulation results, as we would expect given that the lattice spacing, $a$, was computed accurately in our theoretical model (see fig. \ref{lat_derive}).

The cluster radius at each softness changes little with activity, see fig. \ref{phase}b, and remains roughly unchanged with system concentration at high activities ($\mathrm{Pe}>150$), see SI, fig. S11 for $r_\text{c}$ vs $\mathrm{Pe}$ of $\phi=0.45$ and $0.55$. The theoretical predictions of cluster radius given by eq.~\ref{rc} are in good agreement with simulation results (fig.~\ref{phase}b). 

Our analysis of simulations has thus far treated the cluster as a single entity. But as mentioned earlier, it is useful to distinguish two regions within the dense phase: a bulk and an interface. We define the interface as the region, where particles are orientationally aligned (pointing towards the center of mass of the cluster), and the bulk as everywhere else in the dense phase (where there is no orientational alignment), see Fig.5 (a-c). The interface width shows a weak dependence on softness (Fig.5(e)) but is mostly found to be constant over all activities (see SI, fig. S12), area fractions (see SI, fig. S12), and simulation box sizes (see SI, fig. S13) signifying that this measured interface width is an intrinsic quantity of the system. We will approximate the regions as a function of the distance from the center of mass ($||\mathbf{r}||$): bulk $||\mathbf{r}||/r_\text{c}\approx[0,0.8]$ and interface $||\mathbf{r}||/r_\text{c}\approx[0.8,1.0]$.

The bulk maintains a constant average local area fraction, $\bar{\phi}_\text{local}$, approximately equal to that predicted by our theory, $\phi_\text{theory}$ in eq.~\ref{phid}, see fig.~\ref{radial}d-f. Therefore, the area fraction is nearly constant for the majority of the dense phase, supporting the assumption of a constant lattice spacing for our analytical approach.  In addition, the bulk phase exhibits no orientational alignment, $\alpha(r)=\overline{-\hat{\mathbf{p}}(r)\cdot \mathbf{r}}\approx0$, of the body forces, $\hat{\mathbf{p}}$, towards the cluster's center of mass, $\mathbf{r}$, where $r=||\mathbf{r}||$, see fig.~\ref{radial}a-c. Utilizing the virial formulation of pressure, we can calculate the interparticle pressure (eq. \ref{eq:pressure1}) where we see a non-spatially varying pressure experienced throughout the bulk (fig.~\ref{radial}g-i), signifying an equal degree of compression among bulk particles. Similarly, as softness (fig.~\ref{radial}) or activity (fig.~\ref{radial}) increases, so too does the interparticle pressure within the bulk, enabling a greater degree of particle compression (a trend that is captured, through $a$, in our analytical formulation of pressure as well, eq. \ref{eq:pressureVP}). 

However, still within the dense phase cluster, we find a thin surface layer where the local area fraction begins to decrease from the bulk phase area fraction, $\phi_\text{theory}$, until reaching that of the dilute phase, $\phi_\text{g}$, see fig.~\ref{radial}d-f. The decreasing area fraction at the interface results in a drop in the interparticle pressure (as particles are now at distance greater than $a$ from one another, fig.~\ref{radial}g-i).This monotonically decreasing area fraction between two phases resembles the density profiles of typical equilibrium liquid-gas interfaces \cite{Miyazaki1975, Weeks1977a, Chapela1977}, thus we will henceforth label this surface layer as the dense-dilute interface. The body axes of particles becomes aligned within the interface (pointing towards the interior of the dense phase, fig.~\ref{radial}a-c). As the body-axis simply dictates the direction of a particle's active force, we find that the interface exhibits an inwardly aligned active force density (fig.~\ref{radial}a-c), compressing the bulk particles and giving rise to the opposing, outward interparticle pressure from the bulk of the dense phase. We claim this region of both sharply decreasing area fraction and large inwards orientational alignment is a thin dense-dilute interface layer of width $h$ (see SI, fig. S12), motivating us to create a mathematical definition to accurately identify the start ($r_\text{c}-h$) and end ($r_\text{c}$) of the interface (discussed in detail in the SI, see section \S 4). 

\subsection{\S4.2: Surface tension and momentum transport within dense-dilute interface.}

We showed in section \S~\ref{analytical} that the pressure in the dense phase is $\Pi_\text{d}^\text{P} =\frac{4\sqrt{3} F^\text{a}}{a}$. Also, we know that the pressure in the dilute gas phase is negligible compared to the pressure in the dense phase. The transition from dense to dilute phase properties --including pressure, surface tension, and particle alignment-- occur through a thin, dense-dilute interface. 
Force balance dictates that the jump in force per unit area (traction) across the interface must balance against the force induced by the interface itself. Assuming that the interfacial forces are entirely due to surface tension, the equation describing this force balance reduces to
\begin{equation}
    \Delta \mathbf{\hat{F}}_\text{I}=2\gamma \kappa_\text{m} \hat{\mathbf{n}}+\left(\mathbf{I}-\hat{\mathbf{n}}\hat{\mathbf{n}}\right) \cdot \nabla \gamma,  
    \label{eq:surfactension}
\end{equation}
where $\Delta \hat{\mathbf{F}}_\text{I}=\hat{\mathbf{F}}_\text{d}-\hat{\mathbf{F}}_\text{g}$ is the jump in force per unit area across the interface, $\hat{\mathbf{n}}$ is the normal unit vector of the surface pointing outwards  that is separating the dense and dilute phases, $\gamma$ is the surface tension and $\kappa_\text{m}=1/r_\text{c}$ is the mean curvature of the interface. The first and second terms on the right hand side represents the force jumps along the normal and tangential directions of the surface, respectively. Note that the tangential component 
becomes negligible, compared to the normal direction, in our system. 
The ABP model predicts that phase separated domains coarsen with time, ultimately leading to  a single cluster that scales with the dimension of the simulation box, $r_\text{c}=\kappa_\text{m}^{-1} \sim \sqrt{N}$, as derived in eq.~\ref{newRad}. On first examination, one may think of using Young-Laplace equation, which is the form eq.~\ref{eq:surfactension} takes in stationary drops, to determine the surface tension of the active drop as $\gamma_\text{active}=\left(\Pi_\text{d} -\Pi_\text{g}\right) r_\text{c}/2$, (discussed in detail in the SI, see section \S5).
Here, we assume that the gas pressure is negligible and so $\gamma_\text{active}\approx\Pi_\text{d} r_\text{c}/2$.
Substituting eqs.~\ref{eq:pressureVP}-\ref{newRad} to compute a \emph{positive} surface tension at high activities, $\mathrm{Pe} \gg 1$:
$$
    \gamma_\text{active}=4\sqrt{\frac{3\sqrt{3}}{2\pi}}\sqrt{N}F^a.
$$
 The computed surface tension through this formulation is independent of, $\epsilon$, $\phi$ and $\mathrm{Pe}$. Within this formulation, the predicted active surface tension ($\gamma_\text{active}$) is linearly increasing with the dimensions of the simulation box ($\sqrt{N}$) without limit. However, the surface tension should be an intrinsic property of the system and, thus, must be independent of system size. 
 
\begin{figure}[ht!]
\centering
  \includegraphics[width=0.45\textwidth, trim={1.8cm 1.6cm 1.5cm 1.8cm},clip]{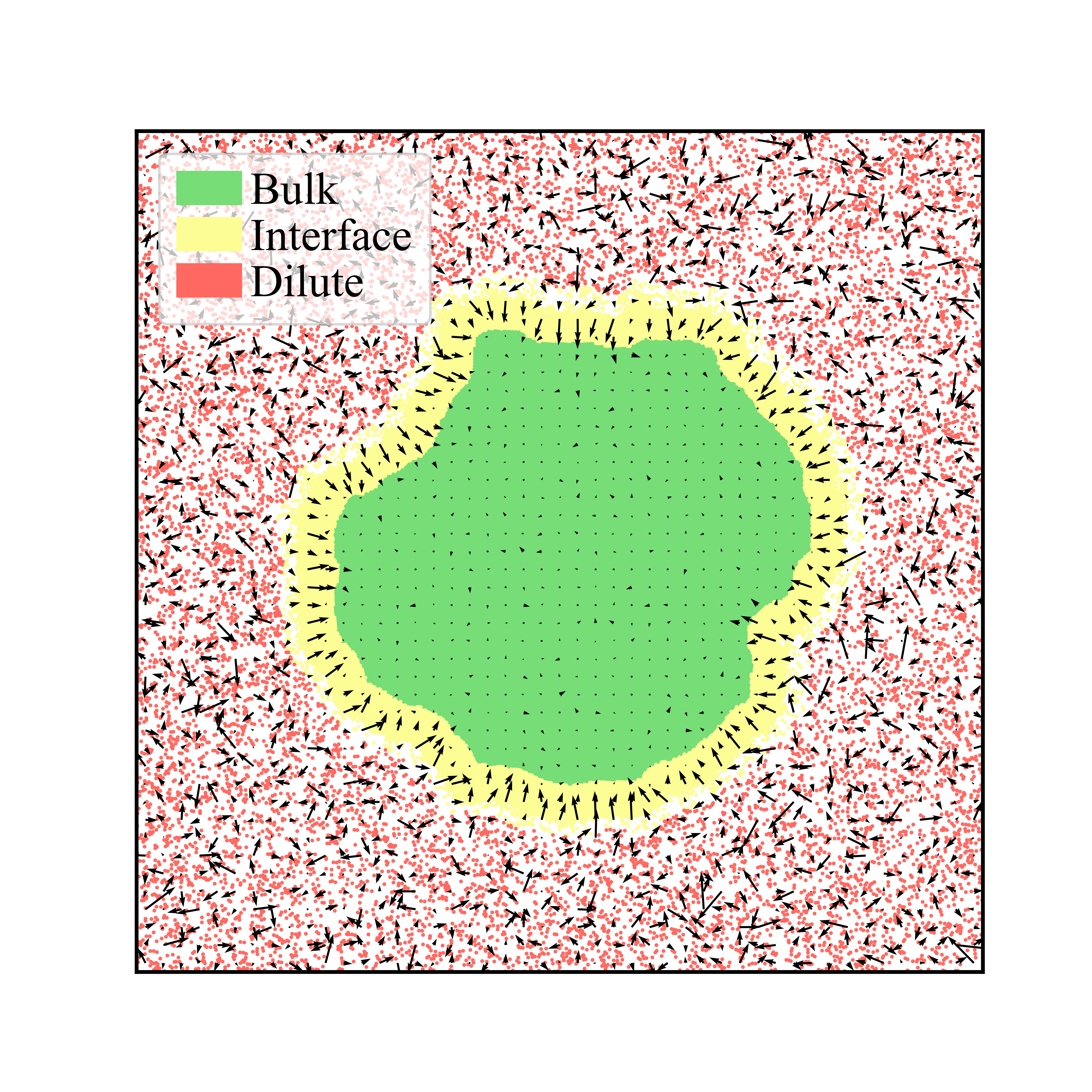}
  \caption{Steady-state ABP system with $\mathrm{Pe}=500$, $\phi=0.55$, and $\epsilon=1.0$ corresponding to a simulation frame at $\tau=186\tau_\text{r}$. The bulk (green), interface (yellow), and dilute (red) phases are labeled according to the average interface width for the system, $h\approx25$. Particles are binned and the average orientation per bin is plotted as the arrows. It is evident that there is essentially zero alignment in the bulk while the interface is highly aligned towards the interior of the cluster.  The orientation of the gas is highly random as particles are freely moving with minimal interactions.  Our observations for the general alignment trends are in agreement with fig.~\ref{radial}a-c.}
  \label{surfacetense}
\end{figure}

 A closer examination of this formulation reveals why it cannot be used to measure surface tension. First, note that in this treatment the surface forces arise from the the net inward orientation of active particles normal to the cluster boundary (see fig.~\ref{surfacetense}), leading to a pressure jump across the surface. This is true whether the interface is flat or curved.  In contrast, in mechanical and thermodynamic formulations of the surface tension for gas-liquid interfaces of passive systems  \cite{Navascues1979}, we have the pressure coexistance condition \textit{i.e}. the pressures in the gas and liquid phase are equal. The surface tension arises due to tangential interfacial forces along the boundary that resist the increase in the surface area \cite{Kirkwood1949}. In other words, the net alignment of particles at the interface of ABP clusters are not acting to minimize surface area; instead they arise in response to normal stress gradients between two phases, including pressure gradients. 
 
To resolve these inconsistencies, we separate this alignment term from the calculations of surface tension and explicitly include it in the momentum equation (eq.~\ref{eq:surfactension}) as a body force, $n(\mathbf{r}) F^\text{a} \hat{\mathbf{p}}(\mathbf{r})$, while still maintaining the term modeling force jump due to surface tension, $\Delta \hat{\mathbf{F}}_\text{I}$. 
Given that the particles within the dense phase move as a rigid body with no relative motion with respect to the fluid, we can neglect the hydrodynamic forces and stresses. In this limit 
the momentum equation in a liquid-gas interface reduces to 
\begin{equation}
    n(\mathbf{r}) F^\text{a} \hat{\mathbf{p}}(\mathbf{r})+\mathbf{f}_\text{I}-\nabla \Pi=\mathbf{0},
\end{equation}\label{momnew}
where $\Pi$ is the macroscopically consistent (true) pressure of the system. 
Note that we have rewritten the interfacial force ($\Delta \hat{\mathbf{F}}_\text{I}$) as a body force in the momentum equation: $\mathbf{f}_\text{I}=\Delta \mathbf{\hat{F}}_\text{I} \delta(\mathbf{x}-\Gamma)$, where $\delta(\mathbf{x}-\Gamma)$ is the Dirac delta function ensuring the surface tension term is localized to the interface region, defined by $\Gamma$. \footnotemark

\footnotetext{Note that while we have presented the interface in the continuum limit as a surface with no volume (line in 2D), in simulations these variations in properties occur over an interface with finite thickness. This is, in practice, similar to approximating Dirac delta function with a smooth and differentiable function with finite, yet small, spreading length, as done in numerical techniques such as immersed boundary method \cite{Peskin2002}. The alignment term $n(\mathbf{r}) F^\text{a} \hat{\mathbf{p}}(\mathbf{r})$ is analogous to the gravitational body force that appear in the formulation of the pendant drop experiment for determining surface tension \cite{Andreas1938}.} 

Assuming that the thickness of the interface is much smaller than the radius of the dense phase, $h/r_\text{c}\ll 1$, and that surface tension is spatially constant, the momentum equation in the radial direction across the interface simplifies to 
\begin{equation}
    -n(r)F^\text{a}\alpha(r)+2\gamma \kappa_\text{m} \delta(r-r_\text{I}) -\frac{\mathrm{d}\Pi}{\mathrm{dr}}=0,
    \label{eq:bl1}
\end{equation}
where $r=0$ and $r=h$ specify the boundaries of the interface residing at the end of the bulk phase and at the cluster radius, respectively, $0<r_\text{I}<h$ is the approximate position of the interface; and $\alpha(r)=\overline{-\hat{\mathbf{p}}(r)\cdot \mathbf{r}}>0$ is the projection of the active force in the radial direction (see fig.~\ref{radial}a-c), signifying the net orientational alignment towards the cluster's center of mass. 

Multiplying both sides of eq.~\ref{eq:bl1} by $\mathrm{dr}$ and integrating across the interface gives an expression for computing surface tension:
\begin{equation}
    \gamma_\text{true}=\frac{1}{2\kappa_\text{m}}\left(\int_0^h n(r)F^\text{a}\alpha(r)\mathrm{dr} -\Pi_\text{d}\right) 
    \label{eq:truesurfacetension}
\end{equation}

\begin{figure}[ht!]
\centering
    \includegraphics[width=0.45\textwidth]{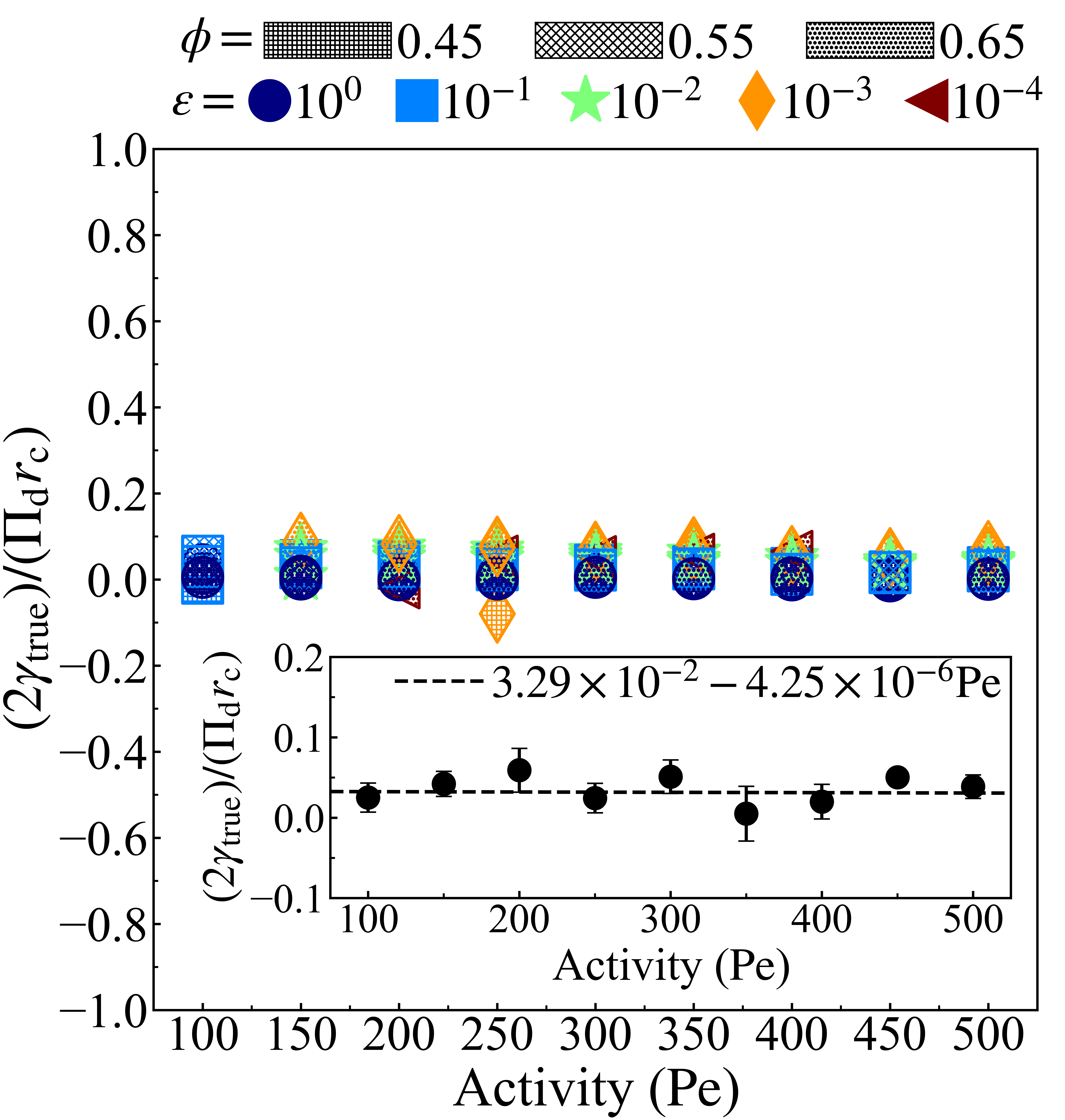}
  \caption{The non-dimensional surface tension, $(2\gamma_\mathrm{true})/(\Pi_\text{d}r_\text{c})$, calculated via eq.~\ref{eq:truesurfacetension} using values for $\alpha(r)$ (fig.~\ref{radial}a-c), $n(r)$ (fig.~\ref{radial}d-f), $r_\text{c}$ (fig.~\ref{phase}b), and $\Pi_\text{d}$ (Hollow circles from fig.~\ref{pressurerad}) measured from simulation.  At all activities, $\gamma_\mathrm{true}$ remains approximately constant near zero with a slight bias in the positive direction (Discussed in detail in the SI, \S 6).  The inset shows the normalized surface tensions averaged over softness ($\epsilon$) and area fraction ($\phi$) at each activity with error bars corresponding to a single standard deviation. In the inset, all surface tension measurements (colored) are fitted (dashed line) such that we do not bias low activity where fewer systems undergo MIPS.  The line of best fit is found to be approximately constant near zero while being encompassed in the standard deviation at most activities.}
  \label{surfacetense_norm}
\end{figure}

\begin{figure}
\centering
  \includegraphics[width=0.45\textwidth, trim={0. 0. 0.1cm 0.3cm},clip]{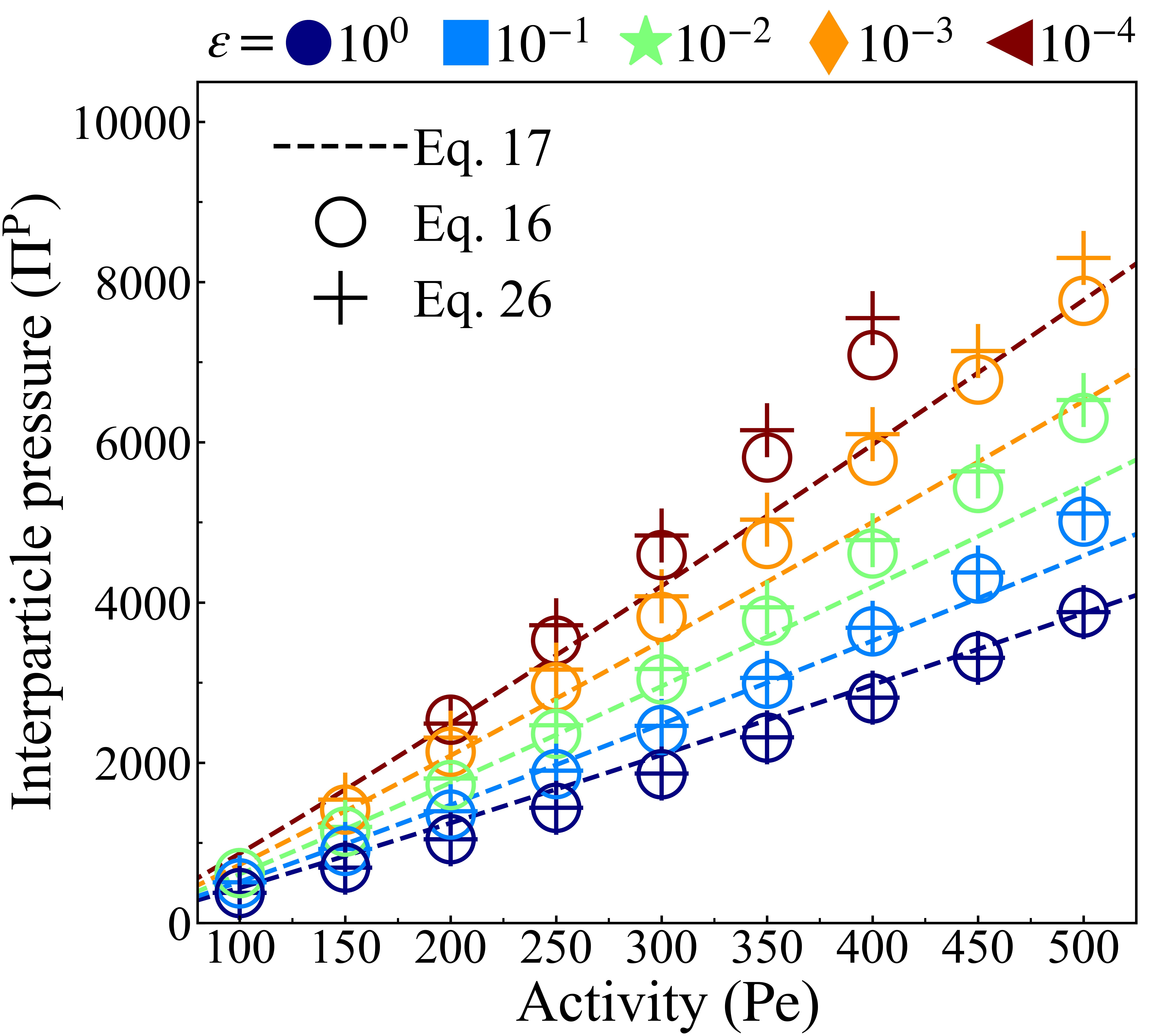}
  \caption{Interparticle pressure computed from the analytical pair-force approach (dashed lines, eq.~\ref{eq:pressureVP}) at distinct particle softness (color) and averaged over the steady state lasting for $\tau\ge50\tau_\text{r}$. Simulation data calculated via the microscopic approach (eq. ~\ref{eq:pressure1}, hollow circles) at $\phi=0.65$ demonstrates good fit for stiff interparticle potential. The quality of fit decays with decreasing stiffness. Simulation data  calculated via the continuum approach (eq.~\ref{presscont}, plus markers) at $\phi=0.65$ show good agreement with both eq.~\ref{eq:pressureVP} and eq.~\ref{eq:pressure1}, demonstrating the possibility to accurately calculate pressure using either a microscopic or a continuum-based approach. In addition, increasing particle activity and softness correspond to smaller clusters (see fig.~\ref{phase}b) with a higher interparticle pressure.}
  \label{pressurerad}
\end{figure}

\noindent Fig. \ref{surfacetense_norm} shows the computed value of surface tension from eq.~\ref{eq:truesurfacetension}, utilizing simulation data for $\alpha(r)$ (fig. \ref{radial}a-c), $n(r)$ (fig. \ref{radial}d-f), $\Pi_\text{d}$ (total interparticle pressure calculated by eq.~\ref{eq:pressureVP} in fig. \ref{pressurerad}), and $h$ (see SI, fig. S12). Note that the surface tension is made dimensionless through dividing by $\gamma_\text{active}=r_\text{c}\Pi_\text{d}/2$.
As it can be seen, the computed surface tension fluctuates around zero without any apparent dependency on softness, activity, and area fraction (discussed in detail in the SI, see section \S 4). This finding is in line with those of Omar \textit{et al.} who also found the surface tension to be nearly zero \cite{Omar2020MicroscopicMatter}. In addition, we find the surface tension is approximately independent of simulation box size (See SI, fig. S13). 

Now that we have established that interfacial forces are negligible compared to gradients of stress normal to the boundary, we can substitute $\gamma_\mathrm{true}\approx0$ into eq. \ref{eq:truesurfacetension} and compute the pressure in the dense phase by evaluating the following integral:

\begin{equation}\label{presscont}
\Pi_\mathrm{d}=F^\text{a}\int_0^h n(r)\alpha(r)\mathrm{dr}
\end{equation} 

\noindent where the values of $\alpha(r)$ (fig.~\ref{radial}a-c) and $n(r)$ (fig.~\ref{radial}d-f) are provided by simulations. 

Having discussed both the virial formulation for calculating the interparticle pressure within the bulk dense phase and the continuum formulation for calculating the pressure arising from the aligned body forces at the interface, we proceed by calculating the total pressure experienced by each region in addition to the resulting pressure equivalence. We start by measuring the total interparticle pressure experienced by each particle in the bulk dense phase (Hollow circles in fig.~\ref{pressurerad}) from its nearest neighbors using the virial formulation of pressure (eq.~\ref{eq:pressure1}).  The total interparticle within the bulk dense phase agrees excellently with our analytical predictions (dashed line in fig.~\ref{pressurerad}), which are linearly increasing with activity and have a slope that increases with softer particles, signifying the greater degree of compression for softer particles in the bulk dense phase. Secondly, using data from fig.~\ref{radial}a-f and eq. \ref{presscont}, we obtain the total pressure from the aligned body-forces in the interface (Plus markers in fig.~\ref{pressurerad}), which we found to be approximately equal to the interparticle pressure of the bulk dense phase (Hollow circles in fig.~\ref{pressurerad}) at every activity and softness. As the pressure of the gas phase is negligible, this finding satisfies a steady-state force balance: the aligned active body-forces at the cluster interface are offset by compressing the particles in the cluster interior to an equilibrium separation, providing an outward interparticle pressure which balances this directed active body-force. 

The agreement of interparticle pressure with the true pressure of the system in the macroscopic scale strongly supports the argument given by Omar et al. that the \emph{swim pressure}, introduced in earlier studies by Takatori and coauthors \cite{Takatori2014a}, should not be included in point-wise definition of the true stress in the continuum scale and the true stress can be computed using the same processes as in passive systems. 
Of course, particle activity does change the stress indirectly through generating a body force due to the net alignment of particles and density gradients across the interface. 
\begin{figure}[ht!]
\centering
    \includegraphics[width=0.45\textwidth]{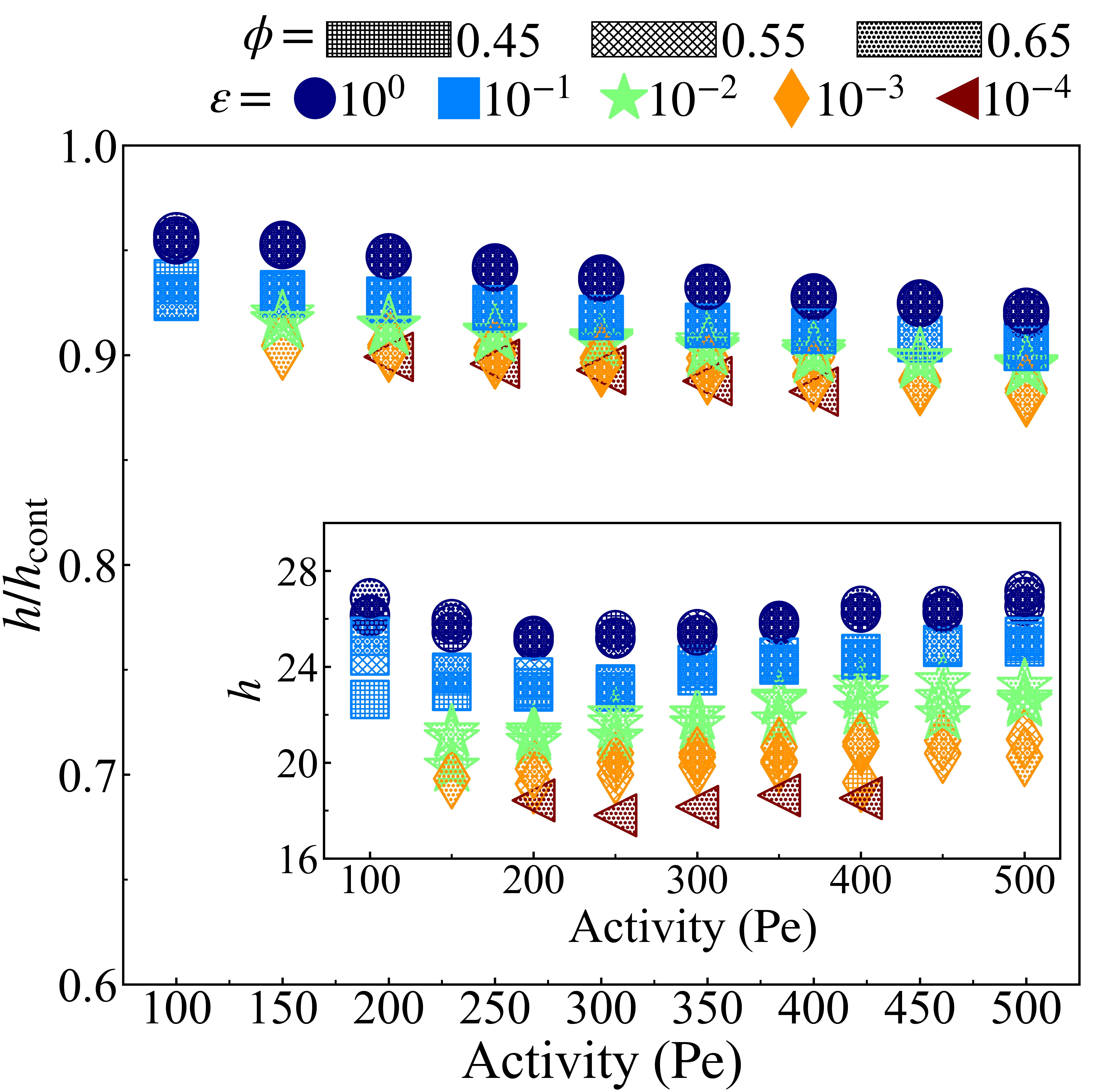}
  \caption{The ratio ($h/h_\text{cont}$) of the interface width ($h$, see SI fig. S12) calculated via the method described in section \S \ref{clustProps} and the interface width ($h_\text{cont}$, see SI fig. S16) calculated through the continuum method (eq.~\ref{simppressint}).  At all activities, $h$ is less than $h_\text{cont}$ by at most $\approx15\%$.  The inset shows the width of the interface measured via simulation ($h$).  When considering the dimensionless interface thickness ($h/a$. See SI, fig. S12), where $a$ decreases with both activity and softness (see fig.~\ref{lat_derive}), the interface width consists of more particles for both more active systems at constant particle stiffness ($\epsilon$) and softer particle systems at constant activity ($\mathrm{Pe}$).  The system area fraction ($\phi$) has a negligible influence on the interface width, similar to its role in the surface tension.}
  \label{intwidth}
\end{figure}

Next, we ask what determines the thickness of the dense-dilute interface. Our simulations at the same $\phi$, $\epsilon$ and $\mathrm{Pe}$ at different simulation box sizes show that, unlike the cluster radius, the interface thickness, $h$, remains unchanged. Previous studies show that the thickness of the boundary layer that forms by accumulation of ABPs near the walls scales inversely with P\'eclet number \cite{Yan2015}. In contrast, the interface thickness in our case is  independent of $\mathrm{Pe}$, as shown in  fig.~\ref{intwidth}.

What, then, determines the interface thickness? Moving radially outwards from the center of mass of the dense phase, the start of the dense-dilute interface is marked by a decrease in the pressure (fig. \ref{radial}g-i) and density (fig. \ref{radial}d-f), and an increase in the alignment, $\alpha$ (fig. \ref{radial}a-c); whereas, the end of the interface is marked by the pressure dropping to nearly zero and $\alpha$ undergoing a 
sharp decrease from its maximum to match the dilute pressure. Following these observations, we rewrite eq.~\ref{eq:truesurfacetension},
using the following change of variables: $$\tilde{\phi}=\phi/\phi_{d},\quad \tilde{\alpha}=\alpha/\alpha_\text{max},\quad \tilde{r}=r/h,$$ 
where $\alpha_\text{max} \approx 0.45$ is the maximum value of $\alpha(r)$  from simulations (see SI, fig. S15). Applying these change of variables, dropping the surface tension contribution and integrating eq.~\ref{eq:truesurfacetension} across the interface gives an expression for the interface thickness in terms of the pressure in the dense phase:  
\begin{equation}\label{simppressint}
h_\text{cont}=\left(\frac{\sqrt{3}}{2\alpha_\text{max}}\right)\left(\frac{\Pi_\text{d}}{F^\text{a}}\right) a^2 \mathsf{I}
\end{equation}
where the term, $h_\mathrm{cont}$ (see SI, fig. S16), denotes 
calculation of the interface thickness using the momentum equation in \emph{continuum} length-scale, and 
$$\mathsf{I}=\left(\int_0^1 \tilde{\phi}(\tilde{r})\tilde{\alpha}(\tilde{r})d\tilde{r} \right)^{-1}.$$ 
Given that the integral term $\mathsf{I}$ only contains scaled variables, $\tilde{\phi}<1$, $\tilde{\alpha}<1$, we expect it to be independent of $\mathrm{Pe}$, $\epsilon$ and $\phi$. The numerical evaluation of 
this integral using simulation results (see SI, fig. S17) confirms this assumption where $\mathsf{I}\approx3.0$ for all $\mathrm{Pe}$, $\epsilon$, and $\phi$. Similar to the surface tension ($\gamma_\mathrm{true}$), the interface width ($h_\mathrm{cont}$) is found to be approximately independent of the simulation box size (see SI, fig. S18). 

Fig. \ref{intwidth} shows the ratio of the interface thickness measured from simulation, $h$ as detailed in section \S3, to the calculated value of interface thickness from eq.~\ref{simppressint}, $h_\text{cont}$, vs $\mathrm{Pe}$ for different values of $\epsilon$ and $\phi$. As can be seen, the ratio remains close to $0.9$ for all values of $\mathrm{Pe}$, $\phi$ and $\epsilon$. The close agreement between the continuum calculations and simulation results is yet another observation in agreement with formulating the pointwise true pressure in the continuum scale as the pressure that arises from interparticle forces and negligible surface forces. 

\section{\S5: Conclusions} \label{conclusions}

A lot of studies have focused solely on the process of phase separation but not the resulting steady-state dense phase. Therefore, in this paper, we characterize and predict the properties of the dense phase itself, such as the area fraction, lattice spacing, and size. To do so, we developed a simple, microscopic analytical approach which relies on (1) the approximation of an HCP dense phase and (2) that each particle interacts with each of its neighbors with an \textit{average} pair force. The microscopic, analytical approach demonstrates reasonable accuracy in reproducing the trends in simulated data for area fraction of the dense and dilute phases and size of the dense phase. These results generalize to ABP systems at any particle softness, activity, simulation box size, or area fraction. Though we utilized the WCA potential to determine interparticle interactions, we fully expect our construction to apply to other short-range repulsive potentials.  An experimental validation of these results is certainly viable. We expect that similar synthetic principles to induce phoresis in hard-sphere colloids, \textit{e.g.} decomposition of oxygen in a hydrogen peroxide solution at the particle surface~\cite{Palacci2013}, can be extended to soft particles, such as polymer functionalized colloids. Alternatively, motility can be induced via polymer chains as is evidenced within cells~\cite{Brangwynne2008CytoplasmicUp} and which has caused the theoretical examination of phoretic polymer chains~\cite{Sarkar2014RingPolymer}.

Though much progress has been made in understanding how this nonequilibrium phase separation gives rise to a dynamic steady-state, one looming question remains: how does the presence of activity influence stress/force generation in the continuum scale and how are these stresses/forces linked to the collective behavior of the system? Many studies have tried to explain this nonequilibrium phenomenon from a thermodynamic perspective via a mechanical equation of state; however, these theories give disagreeing results for important physical properties, such as surface tension.  The main difference between these studies is whether activity gives rise to a stress that acts as either a spatially uniform state variable (the \textit{swim pressure} \cite{Takatori2014a, Fily2014, Takatori2017, Mallory2020}) or a spatially varying body force density \cite{Epstein2019, Omar2020MicroscopicMatter}. In the former group of studies the stress is defined as a volume-averaged quantity within the container such that there are no spatial gradients, and it is shown, through theory and simulation, that the activity induces an extra term referred to as \emph{swim pressure} \cite{Takatori2014a}. Upon utilizing the swim pressure to describe the system, one can accurately predict many emergent, macroscopic properties, such as determining the onset of MIPS \cite{Winkler2015, Levis2017} and explaining how active pressure being non-monotonic with activity and area fraction gives rise to a phase transition \cite{Takatori2014a, Winkler2015, Levis2017} .

However, problems arise when we seek answers to localized phenomena.  Omar \textit{et al.} recently showed that 
including the swim pressure in the description of total pressure results in extremely negative values of surface tension \cite{Bialke2015a, MariniBettoloMarconi2015, Paliwal2017, Patch2018, Solon2018b}, in contrast to passive systems. Extension of the statistical mechanics derived for passive systems to its active counterparts is reliant on the system being homogeneous with no concentration or alignment gradients.   However, our active systems demonstrate a monotonically decreasing area fraction at the highly aligned dense-dilute interface, which gives rise to this negative surface tension term when treating the volume-averaged \textit{swim pressure} as the active analogue to the osmotic pressure. Though a volume-averaged treatment of pressure fails to explain surface tension, it does work on a macroscopic scale where the volume-averaged body forces cancel out, giving rise to the \textit{swim pressure} in the momentum equation. 

Therefore, accurate characterization of localized phenomena, like surface tension, requires a point-wise treatment of pressure as we cannot define a swim pressure at an interface where there is no volume-averaging involved. As a result, we no longer treat activity's contribution to pressure as a spatially independent variable but instead as a spatially varying body force. Activity produces aligned body forces acting on the boundary around the container or between phases, not a stress, which is reserved for that as defined in a passive system in order for our definition to \textit{always} be correct. By \textit{always}, we mean that though a volume-averaged approach (utilized commonly in traditional, equilibrium thermodynamics) works as an exception \cite{Solon2015a} for determining macroscopic properties in our ABP systems, a point-wise approach to pressure, specifically by treating activity as a body force to account for gradients in density and alignment at surfaces, satisfies a mechanical force balance on both the microscopic and macroscopic scale, enabling us to explain and predict emergent behavior. Omar \textit{et al.} showed that if only the stress due to passive forces in ABP systems are considered in the definition of pressure while activity is treated separately as a spatially varying body-force density, the predicted surface tension, which relies on gradients of properties across the interface \cite{Navascues1979}, becomes negligible. 

The results presented here confirm that of Omar \textit{et al.}, demonstrating that the point-wise mechanical effect of activity is to generate the gradients in concentration and alignment of particles, resulting in a body force (not stress), $\alpha(r) n(r) F^\text{a}$, in the continuum level. Using the virial formulation of pressure, we have derived an analytical expression for the interparticle pressure in the bulk of the dense phase that agrees strongly with simulations. We also derived a second, continuum approach that utilizes the radial alignment, radial area fraction, and dense phase pressure from simulation to calculate the stress from aligned body forces at the interface, which approximately equals the interparticle pressure of the dense phase. As a result, the surface tension is approximately equal to zero for all softnesses, activities, area fractions, and simulation box size, demonstrating how the surface tension is an intrinsic property of the system. As such, we similarly confirmed that the interface width was an intrinsic quantity of the system, as similarly predicted by a local free-energy approach in equilibrium liquid-vapour interfaces \cite{vanderwaal1979, Cahn1958, Fisk1969, Singh1977}, with both the analytical interface width and that measured via simulation agreeing within $10\%$ for all activities, softnesses, area fractions, and system box sizes.

While our results demonstrate the complex behavior that is accessible to monodisperse active mixtures of varying stiffness, the work presented here is only the first step. A number of interesting future directions are evident that will help us further understand the mechanism behind nonequilibrium steady states, namely those characterizing the dense-dilute interface. Although we found surface tension plays a negligible role in mechanically maintaining the steady state, it could play a role in other important physical properties, such as controlling fluctuations and particle flows at the dense phase surface. 

If the interface is disturbed by an external force, there will be a local displacement of interfacial particles in the immediate vicinity that continues to travel tangentially across the interface while decaying in the process, like a wave \cite{Wertheim1976}. These surface fluctuations give rise to long-range correlations in density across the interface, which are consistent with a description of the surface in terms of capillary waves that are thermally excited against surface tension or an external force \cite{Evans1979}, enabling us to characterize the fluctuations similarly \cite{Wysocki2016} and understand their role in stability, like cascading, avalanche events \cite{Redner2013, Stenhammar2015}. Surface tension could also mitigate long-term surface instabilities through surface flows in ABP systems as seen in other liquid-gas interfaces \cite{Keller1983}. In ABP systems, curvature-dependent surface tension drives sustained local tangential motion of particles on either side of the interface, suggesting a redirection of particles to heal local fluctuations and promote stability \cite{Patch2018}, potentially maintaining the aligned body forces at interface that stabilizes the cluster. In addition, many other interfacial properties of ABP systems have been connected to that of equilibrium phases \cite{MariniBettoloMarconi2016, Speck2016, Prymidis2016, Lee2017, Solon2018, Tjhung2018}, necessitating deeper study of the interface and surface tension's role in mechanical stability of non-equilibrium steady states.

\section*{Conflicts of interest}
There are no conflicts to declare

\section*{Acknowledgements}
Nicholas Lauersdorf and Thomas Kolb contributed equally to this work. This material is based upon work supported by the National Science Foundation Graduate Research Fellowship under Grant No. (NSF grant number DGE-1650116).




\bibliography{Mono_soft} 

\end{document}